\documentclass{ieeeaccess}
\usepackage{cite}
\usepackage{amsmath,amssymb,amsfonts}
\usepackage{algorithmic}
\usepackage{graphicx}
\usepackage{textcomp}
\def\BibTeX{{\rm B\kern-.05em{\sc i\kern-.025em b}\kern-.08em
    T\kern-.1667em\lower.7ex\hbox{E}\kern-.125emX}}
\begin{document}
\history{Date of publication xxxx 10, 2021, date of current version xxxx 10, 2021.}
\doi{10.1109/ACCESS.2017.DOI}

\title{
New data poison attacks on machine learning classifiers for mobile exfiltration
}
\author{
{Miguel A. Ramirez}\authorrefmark{1},
{Sangyoung Yoon}\authorrefmark{1},
{Ernesto Damiani}\authorrefmark{1}, 
{Hussam Al Hamadi}\authorrefmark{1}, 
{Claudio Agostino Ardagna}\authorrefmark{2}, 
{Nicola Bena}\authorrefmark{2}, 
{Young-Ji Byon}\authorrefmark{3}, 
{Tae-Yeon Kim}\authorrefmark{3}, 
{Chung-Suk Cho}\authorrefmark{3}, 
and {Chan Yeob Yeun}\authorrefmark{1}
}

%


\address[1]{Center for Cyber-Physical Systems, EECS Department, Khalifa University of Science and Technology, Abu Dhabi, United Arab Emirates}
\address[2]{Dipartimento di Informatica, Universita degli Studi di Milano, Milano, Italy}
\address[3]{Department of Civil Infrastructure and Environmental Engineering, Khalifa University of Science and Technology, Abu Dhabi, UAE}

\tfootnote{This work was supported in part by Technology Innovation Institute (TII) under Grant 8424000394.}

\markboth
{Author \headeretal: Preparation of Papers for IEEE TRANSACTIONS and JOURNALS}
{Author \headeretal: Preparation of Papers for IEEE TRANSACTIONS and JOURNALS}

\corresp{Corresponding author: Chan Yeob Yeun (e-mail: chan.yeun@ku.ac.ae).}

\begin{abstract}
Most recent studies have shown several vulnerabilities to attacks with the potential to jeopardize the integrity of the model, opening in a few recent years a new window of opportunity in terms of cyber-security. The main interest of this paper is directed towards data poisoning attacks involving label-flipping, this kind of attacks occur during the training phase, being the aim of the attacker to compromise the integrity of the targeted machine learning model by drastically reducing the overall accuracy of the model and/or achieving the missclassification of determined samples. This paper is conducted with intention of proposing two new kinds of data poisoning attacks based on label-flipping, the targeted of the attack is represented by a variety of machine learning classifiers dedicated for malware detection using mobile exfiltration data. With that, the proposed attacks are proven to be model-agnostic, having successfully corrupted a wide variety of machine learning models; Logistic Regression, Decision Tree, Random Forest and KNN are some examples. The first attack is performs label-flipping actions randomly while the second attacks performs label flipping only one of the 2 classes in particular. The effects of each attack are analyzed in further detail with special emphasis on the accuracy drop and the misclassification rate. Finally, this paper pursuits further research direction by suggesting the development of a defense technique that could promise a feasible detection and/or mitigation mechanisms; such technique should be capable of conferring a certain level of robustness to a target model against potential attackers. 

\end{abstract}


\titlepgskip=-15pt

\maketitle

\section{Introduction}
\label{sec:introduction}
Over the last couple of years, machine learning (ML) models demand the deployment of additional techniques in order to address security related factors since new vulnerabilities are being discovered and could pose a threat to the integrity of the ML model being the target of an attacker \cite{R04}. An attacker could exploit such vulnerabilities causing a negative impact on the performance of the ML model. It is been proven plausible to maliciously compromise the training data in order to affecting the model decision-making process which eventually causes a utter malfunction during testing (or inference) phase.
\\
The need of public and available data is continuously on demand by plenty ML models. A clear example can be seen in smart city systems wherein large amounts of data are gathered by numerous sensors, such as smartphones. Then it comes without saying that the consequences of an attack targeting smart city systems could be devastating and very feasible due to the system heavily dependence over public data. The intention of this paper is directed towards gathering the most representative attack and defense approaches around data poisoning \cite{Paudice_1}. Data poisoning (DP) attacks aim to compromise the integrity of a target model by performing alterations to the required dataset used by the model during the training phase. This causes the model to misclassify samples during the testing phase, representing then a significant reduction in the overall accuracy. 
\\
For all the reasons previously mentioned, there is the urge to develop more advanced defense mechanisms, aiming to enhance robustness of the model to fight back potential DP attacks occurring while training, the further understanding of such vulnerabilities could offer promising results directed towards the development of a defense mechanism capable of detecting and even mitigating the effects of the poisoning data, making by then ML classifiers with a higher resiliency than current ones. This last point has been one of the main center of focus when it comes to malware classification, application of which the target ML models in this work will address specially.
\\
The introduction of a new approach that attains a certain level of immunization against DP in the fields of malware detection involving mobile ex filtration data \cite{Martina}, being this a common topic of interest regarding smart devices and smart cities environments. The main contributions of this paper are listed as follows:
\\
\begin{itemize}
\item This paper covers diverse topics around machine learning security fundamentals, assumptions of attack and defense scenarios, types of data manipulation and vulnerabilities.
\item Related work on potential threats involving data poisoning attacks and related defense mechanisms are showcased, particularly towards manipulation of mislabeled training data such as label-flipping techniques.
\item The methodology entailing crafting a data poisoning attack based on label-flipping, targeting a malware ML classifier. A further evaluation of the effects of the poisoning attack on different ML models, comparing the results obtained with each other, quantifying the effects of the attack and vulnerabilities of each model.
\item Open problems and future research work towards the field of machine learning security is discussed with special emphasis in analyzing more complex label-flipping attack scenarios and other defense mechanism oriented towards detection and mitigation against DP attacks are examined as well.
\end{itemize}

This work is organized as follows: Knowledge background information is explained along with related work on various types of data poisoning attacks and defense mechanisms on Section II. The methodology explaining both the development and evaluation of the proposed attack on target ML models, as well as for the proposed defense mechanism against the proposed attack can be found on Section III. Future work and research directions related to this work can be found on Section IV. Finally, the conclusion and final remarks are shown in Section V.

\section{KNOWLEDGE BACKGROUND}
n this section, an overview of the properties around attacks and defenses is analyzed. The fundamentals of relevant topics involving security in machine learning models are discussed, mainly centered in the assumptions of the attacker and different types of attacks throughout the machine learning life-cycle.

\subsection{Manipulation}
Training data manipulation \cite{A02} is one of kinds of DP attacks by corrupting (or poisoning) the training data during the training phase with the aim to utterly jeopardize the integrity of the ML classifier having trained a wrong classifier, examples of techniques often used by attackers are the modification of data labels, injection of malicious samples and manipulation of the training data. As a result, the overall damage to the target ML model can only be appreciated at the inference phase, having the accuracy of the model drastically reduced. This effect is commonly referred to as accuracy degradation.
\\

Input manipulation is triggering a machine learning system to malfunction by altering the input that is fed into the system \cite{A03}. It would be in the form of an altered image by adding noise or another input that causes the classifier to perform a wrong prediction. Adversarial attacks \cite{Miller_Survey}–\cite{R12} take place during the inference phase, having a ML model already trained and as a result any prediction from the model is considered of high confidence \cite{R09}. Depending on the goal of the attacker an adversarial attack can fall in one of two categories. Referring to a Targeted attack when the input in the form of crafted adversarial examples lead to the target model to misclassify the samples into a specific class defined by the attacker \cite{D_Meng}, \cite{R01}. In contrast, in a Non- targeted attack the crafted adversarial examples aim to cause the target model to misclassify. Nonetheless, there is no need nor interest from the attacker to misclassify into a particular class apart from the correct one. Evasion attacks are also another kind of input manipulation and are different from adversarial attacks in the sense that evasion attacks do not require any knowledge over the training data \cite{R05}.

\subsection{Assumptions of attack-defense}
Security threats to machine learning models are divided into data poisoning (DP) attacks and adversarial attacks, acting the former during the training phase and the later during testing, this difference is shown in Figure \ref{DP}. For the purposes of this work, data poisoning attacks will remain as the main topic of interest.

\Figure[!h][width=0.4\textwidth]{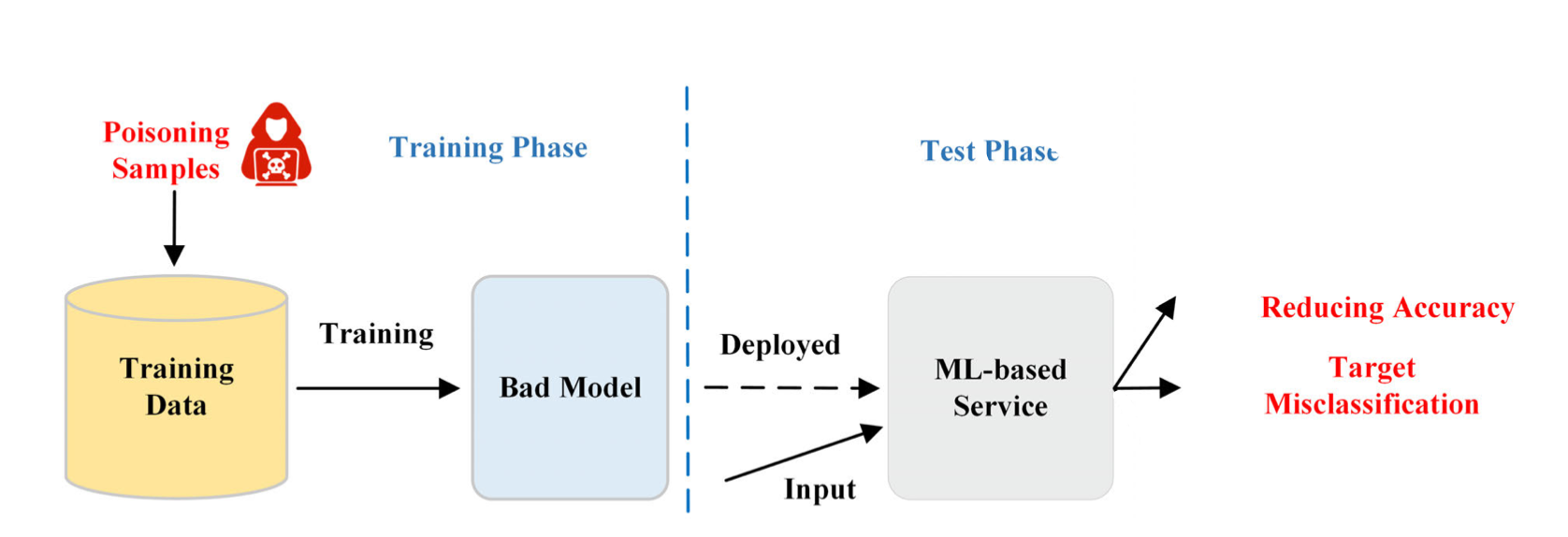}{Data poisoning attacks during training phase affecting testing phase \cite{Liu_Ximeng}. \label{DP}}

One of the most common schemes of the attacker is injecting malicious samples into the target model’s training set, corrupting either the feature values or labels of the training samples. Affect the ML model boundaries by causing significant deviations to a point where the model’s reliability is completely devastated, thus leaving the model susceptible to make wrong predictions. Then the goal of the attacker then is to disrupt the training process aiming to significantly reduce the performance of the target model, causing an accuracy- drop; also generating increasing misclassification rate of the samples during testing.
\\
Assumptions of attackers refer to the prior knowledge (implicit or explicit) over the target model of interest of the attacker, entailing the resources available to the attacker. When conducting experiments, the devised attack is meant to be evaluated against a defense, both attack and defense state assumptions must be declared assuring the conditions that guarantee either scheme’s effectively (eg. attaining the defense to defeat an attack, or viceversa). However, various DP attacks have shown to be successful in spite of having very little knowledge of the target model. An example of this is described in the work \cite{Lowd}, directing a DP attack scheme to naive Bayes email-spam filters by simply sending ‘ham- like’ emails using a black-listed IP address as the sender, then being threatened and labeled as spam, nonetheless these corrupt data will be inevitably used by the spam filter for further training.
\\
The assessment of the influence of the attacker over the training data is commonly defined as the attacker’s capability. Being the primary interest of the attacker is to alter either the feature values or labels as part of the training set. Nevertheless, the attacker is restricted to poison a limited number of samples, commonly referred as a ratio of less than 30\% of the total data samples. More optimized poisoning algorithms have been in constant development during the last decade, aiming to maximize the accuracy degradation and minimize the number of poisoning samples needed to perform the attack.

\subsection{Metrics of interest}
The attack success related to DP attacks is estimated based on the amount of degradation shown by the target model performed during the testing phase. This can be further appreciated once computing the decision matrix, observing the overall misclassification in each class displaying: true positive, false positive, true negative and false negative. Moreover, the effectiveness of the attack is shown in the form a significant drop in the overall accuracy, this is referred as accuracy degradation. 
\\
The employment of additional metrics besides measuring the accuracy have been proposed to reflect and analyze in further detail the overall performance of the target model and by then make comparisons to other model performance. Various metrics for artificial intelligence have been proposed by multiple standard bodies including the International Organization for Standardization (ISO) \cite{A21} and the National Institute of Standards and Technology (NIST) [22]. The metrics for AI typically includes the accuracy, the precision, the recall, the Receiver Operating Characteristic (ROC) and its area1 [23].
\\
Other approaches such as the one from Biggio [24] introduces the security evaluation curves as a way to characterize the performance of a ML model against an intended attack considering various level of knowledge from the attackers side. Thus this approach accomplishes a comprehensive evaluation of the overall security of the model; and by doing so, enables another means to compare assorted defense techniques.

\subsection{ADVERSARIAL CAPABILITIES}
In the testing phase, the attacker naturally will aim to attain further knowledge over the target model in order to increase the effectiveness of adversarial attacks, this can be in the form of any of the following five factors: Feature space, classifier type (e.g. DNN or SVM), classifier learning algorithm, classifier learning hyperparameters and the training dataset.
\\
\begin{itemize}
\item A white-box assumption is commonly defined as an scenario in which the attacker does have complete knowledge over all the five elements already described previously, as well as any defense mechanism already set on top of the model [25]–[27].
\item A black-box assumption is the opposite to white-box assumption, when no knowledge of the target model, albeit query it can be plausible. Nonetheless, it is important to remark that, just having access to the training data grants the upper hand to the attacker over any defender, representing this training data the unadulterated or ‘clean’ dataset, in question [28]–[33].
\item A gray-box assumption is often referred as a middle ground between white-box and black-box scenarios, where the prior knowledge on the attacker’s side can include the feature space, the target classifier; this includes the model architecture, model parameters and the training dataset; however, the defense mechanism on top is unknown to the attacker. The gray-box setting usually is used to evaluate the defense against the adversarial attack [17].
\end{itemize}

\subsection{Explainable Artificial Intelligence}
Explainable Artificial Intelligence (XAI) is a concept that looks for the development of artificial intelligence models with a further improved understanding regarding the main aspects of their decision-making process, this is for us humans to understand why and how a model approaches a prediction. The aim of explainable AI is to develop more reliable strategies that will endow models with a higher level of transparency while retaining high-performance levels, mainly in the form of accuracy \cite{Nicola}.
\\

XAI is extremely necessary nowadays in various models since explainability is often translated into an adequate level of trust in the predictions. Regardless of high performance, a model's predictions cannot always be taken as ground-truth, especially in applications of critical importance in terms of reliability, such as with cybersecurity as well as other future-generation AI partners. In the field of cybersecurity transparency is a must, the lack of it represents dangers. Therefore it has become essential to maintain a good balance between explainability and performance, looking further into the trade-offs they both confer in the field of cybersecurity and the newly introduced technologies around it \cite{Chan}.
\\

Decision trees represent one example of a good approach towards high explainability since it confers higher transparency than other ML models, enabling the user to understand the decision-making progress more easily. Nonetheless, deep learning models perform better than decision tree models, it results to be the algorithm with less explainability.

\subsection{Related Work: Label-flipping attacks}
The most common way to generate this kind of poisoning is by maliciously tampering the labels in the data \cite{X_Liu}, this can be easily achieved by just flipping labels, thus generating mislabeled data, this is shown in \ref{LF}.

\Figure[!h][width=0.3\textwidth]{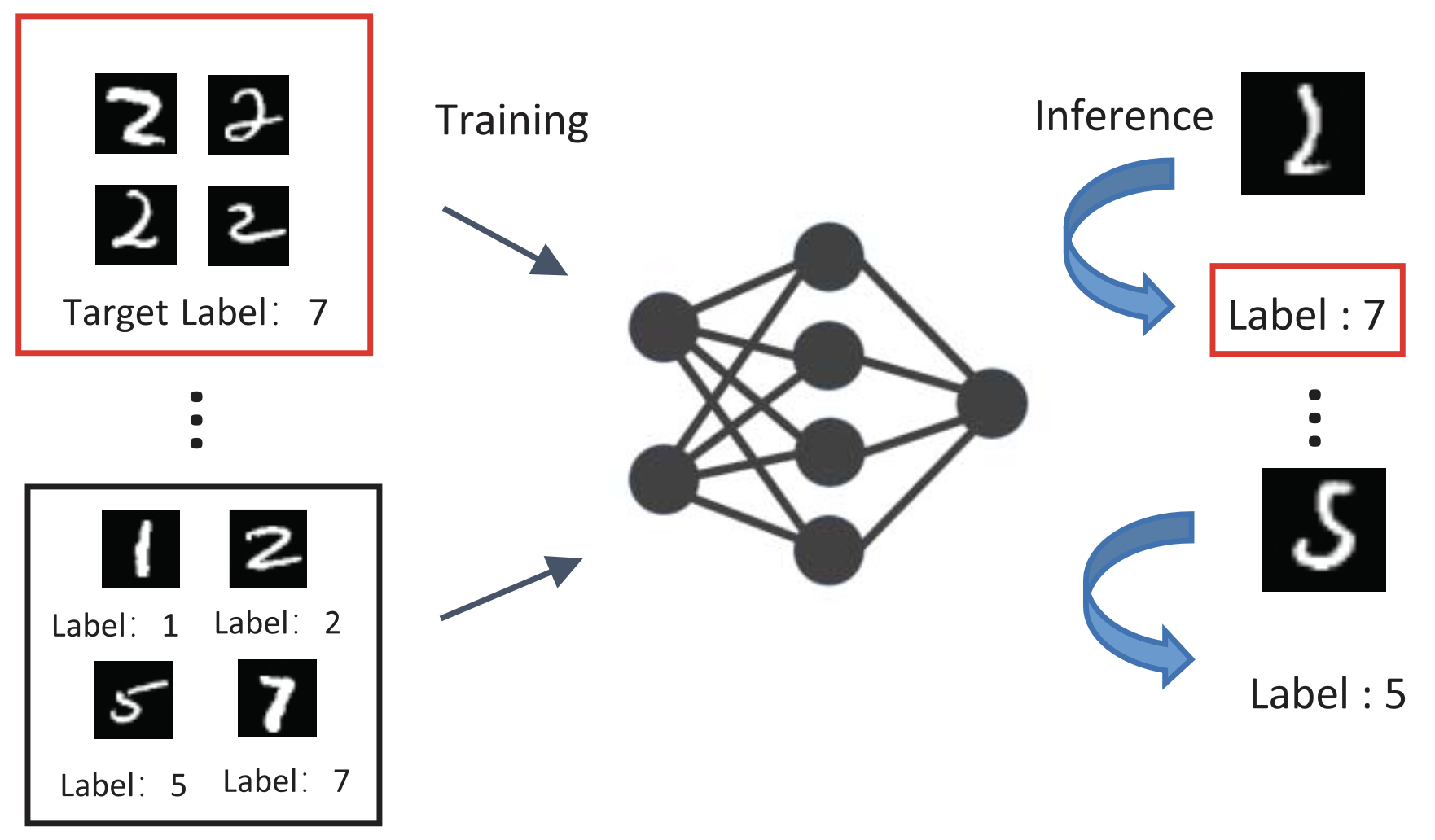}{Misclassification error caused by label-flipping \cite{X_Liu}.\label{LF}}

Label flipping can be performed either randomly or specifically depending on the aims of the attacker; the former aims to reduce the overall accuracy of all classes, the later does not aim to perform significant accuracy reduction, rather it is focus on the misclassification of a determined class in particular.
\\
In the following paragraphs several examples of data poisoning attacks on different types of models are discussed. The center of focus is directed towards data poisoning attacks performed during the training phase involving label-flipping techniques against ML classifiers exclusively.
\\
Paudice et al. \cite{Paudice_2} proposes an optimal label flipping poisoning attacks compromising machine learning classifiers. Label flipping actions are performed following an optimization formulation focused on maximizing the loss function of the target model. This approach is considered computationally intractable due to the inclusion of heuristic functions enabling the label flipping attacks to downscale the computational cost.
\\
The applications of this approach limits itself to binary classification problems and the assumptions of the attack involves complete knowledge over the learning algorithm, loss function, training data and also the set of features used by the ML classifier, turning it basically into an attack on a White-box model. Albeit the list of assumptions appeal to unrealistic scenarios, the analysis emphasizes on worst- case scenarios. The effectiveness of the propose method is demonstrated in three datasets from UCI repository: MNIST, Spambase and BreastCancer; succeeding in increasing the classification error by a factor of 6.0, 4.5 and 2.8, respectively [35].
\\
Xiao et al. [36] reports a successful attack on a SVM model after performing label flipping using an optimized framework capable of procure the label flips which maximizes potential classification errors, causing a significant reduction in the overall accuracy of the classifier. As a potential drawback, this technique naturally implies a high computational over- head as a main requirement.
\\
Federated Learning (FL) in recent years has become relevant in privacy-preserving applications. This is possible since the data gathered from each device or worker is kept locally stored in each device [64], then enabling the training process of a sub-machine learning model individually. As a second step, only the resulting gradients obtained after training are exchanged to a centralized server instead of the raw data, then the centralized server performs the entire training life-cycle by multiple iterations until attaining a desirable accuracy. Due to the nature of FL, malicious users could perform a label- flipping attack [39] by deliberately inserting crafted gradients leading to classification errors during the test phase. In the past it is been proven that a single poisoner can undermine the whole training process and as a result the integrity of the model. Therefore, a robust FL model needs to regard on concerns related not only to data privacy, but also rely on a certain degree of resilience against poisoning attacks and data manipulations. 

\subsection{Related Work: Defenses}
k-Nearest-Neighbours defense scheme [35] is designed to detect malicious data and counteract the effects of the same, being this defense referred as Label sanitization (LS). Label sanitization (LS) bases its defense on the decision boundary of SVM, observing the remoteness of the poisoned samples, commending these samples to be re-labelled. Steihhardt et al. [58] proposes a nearest-neighbor-based mechanism to detect outliers and SVM optimization right after- wards, getting as a result a domain-dependent upper bound associated to the estimated highest drop in accuracy due to a DP attack. A special assumption is made for this scenario, declaring the removal of non-attack outliers inconsequential to the performance of the target model.

Liu et al. [34] addresses the privacy/defense related issue in FL models by showcasing a novel framework called privacy-enhanced FL (PEFL). PELF grants the central server the ability to detect malicious gradients and block poisoner workers. By comparing the malicious gradients, submitted by the poisoner workers, as a set of parameters to the same ones belonging to the honest workers; the difference between malign and benign gradient vectors can be evaluated by calculating the Pearson correlation coefficient [65]. Abnormality behavior is related to a lower correlation coefficient, then the action of the defense mechanism consists on simply setting the weights of the malign model to zero. 
\\
PEFL claims superiority among other similar systems such as Trimmed Mean [64], Krum [66] and Bulyan [67]. Since the proposed scheme does not assume to have any knowledge of the total number of poisoners, posing then a more appropriate defense more suitable for real-case scenarios. Furthermore, PEFL poses a higher resilience against accuracy drops compared to Bulyan and Trimmed Mean due to weight adjustment performed on each gradient which guarantees trustworthiness within the remaining parameters. Observing in the end a maximum attack success rate of 0.04, evidence of the robustness of the model against label- flipping.
\\
SVM Resistance Enhancement [68] is targeted to avoide label-flipping attacks, being SVM particularly vulnerable against this kind of attacks, causing total misclassification due to the computation of erroneous decision boundaries. Thinking ahead about the effects of suspicious data points within the SVM decision boundary, the proposed approach considers a weighted SVM accompanied by KLID (K-LID- SVM). This work introduces K-LID, a new approximation of Local Intrinsic Dimensionality (LID), metric associated to the outliners in data samples. K-LID computation relies on the kernel distance involved in the LID calculation, allowing LID to be computed in high dimensional transformed spaces. Obtaining by such means the LID values and discovering as a result three specific label dependent variations of K-LID capable of counter the effects of label-flipping.
\\
K-LID-SVM attains higher overall stability against five different label- flipping attack variants: Adversarial Label Flip Attack (alfa) attack, ALFA based on Hyperplane Tilting (alfa-tilt), Farfirst, Nearestl and Random label flipping; using five different real-world datasets for a benchmark test: Acoustic, Ijcnn1, Seismic and Splice and MNIST. The defense system attains a drop of 10\% on average in misclassification error rates, this method can distinguish poison samples from honest samples and then suppress the poisoning effect. Therefore, it succeeds in decreasing the potential magnitude of the attack significantly and demonstrating a superior performance than traditional LID-SVM

\Figure[!h][width=0.3\textwidth]{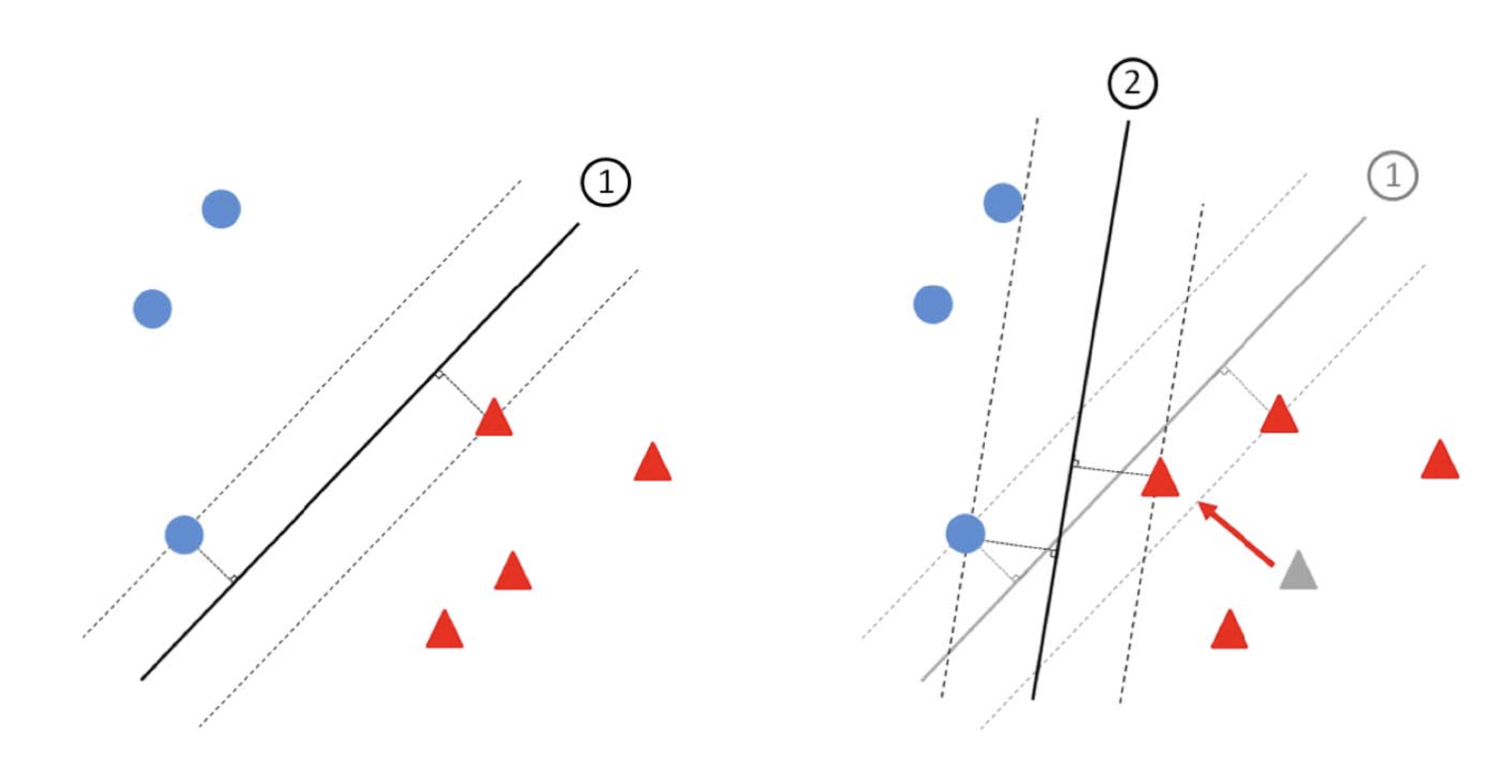}{(Left) SVM classifier decision boundary for two classes (Right) Impact on decision caused by the re-location of one single data point \cite{D_Miller}. \label{SVM}}

In this chapter we will propose an attack strategy able to effectively compromise the integrity of more than one type of ML classifier, causing a significant drop in accuracy and miss classification in one or various classes.

\subsection{Mobile malware classifier}
Morcos et al. \cite{Martina} develops a mobile exfiltration data. This dataset is considered the main data of interest in this work as it represents the data used to feed the ML model comes from an mobile exfiltration detection engine for Android phones. From any application in executions system calls can be observed and studied as partial information on onboard activities, as seen in \ref{execution_program}.

\Figure[!h][width=0.45\textwidth]{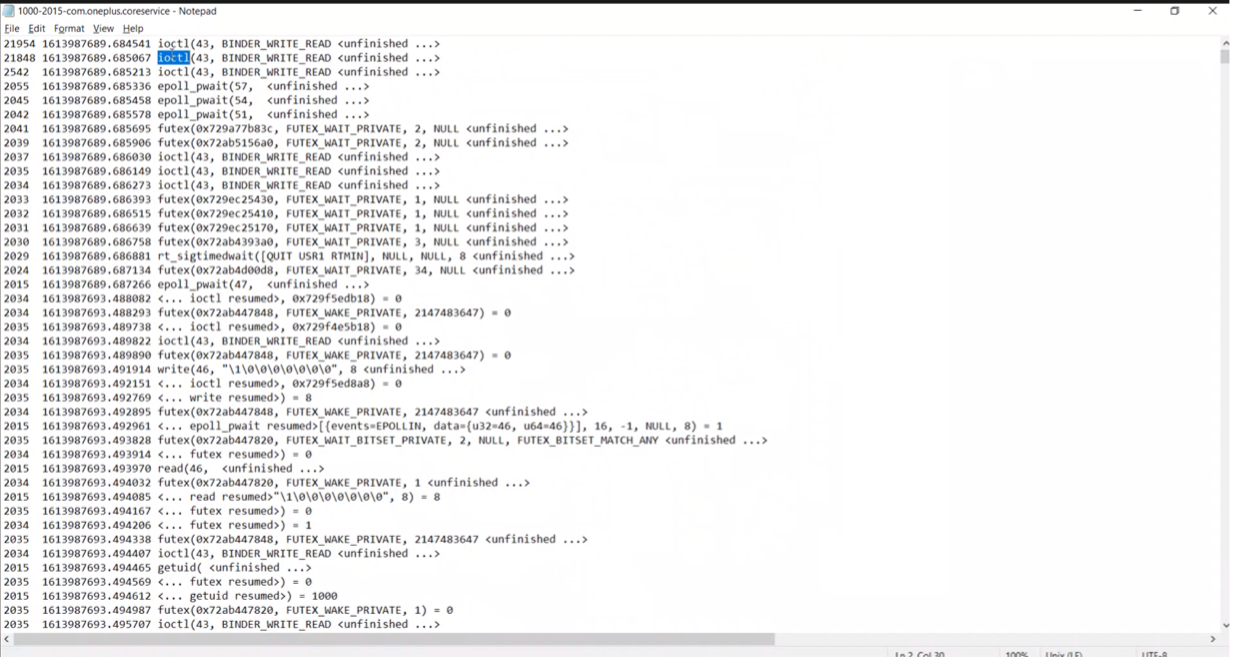}{System calls of program in execution. \label{execution_program}}

Based on this onboard activities system calls can be extracted, then the statistical behavior of a potential malware can be appreciated by the frequency in which determined system calls appear during execution, forming an histogram. The overall approach is described in five steps:
\\
\begin{itemize}
\item Defining app and system-call activity data segmentation and mapping to activity representation.
\item Identifying normal behavior for a given configuration.
\item Identifying exfiltration behavior from sandboxing.
\item Training stateful ML models.
\item Deploying and comparing.
\end{itemize}

\hfill

The feature engineering process entails then selecting the most representative system calls for this malware detection as the features of interest, this process receives the name of Monolithic approach. Once setting this representation this data can be feed into an ML model classifier in further steps, then a ML model can generalize well among the malware data and detect when exfiltration or data leakage occurs.
\\

Another approach that has resulted more accurate for malware detection is by declaring the features as 3-gram of system calls. This is referred as the 3-gram approach, here each feature refer to a specific sequence of three system calls; therefore only after having recorded the set of 3 system calls in the declared order a value can be set. The 3-gram approach serves as the main and only data representation of interest in this work.
\\

In the 3-trigram approach, different variations of the gathered application data have been created by performing a clustering process at different levels of the process tree. In the Figure \ref{clustering} it can be appreciated the process tree for a remote app, the clustering takes place in three levels: Level 0, Level 1 and Level 2.

\Figure[!h][width=0.8\textwidth]{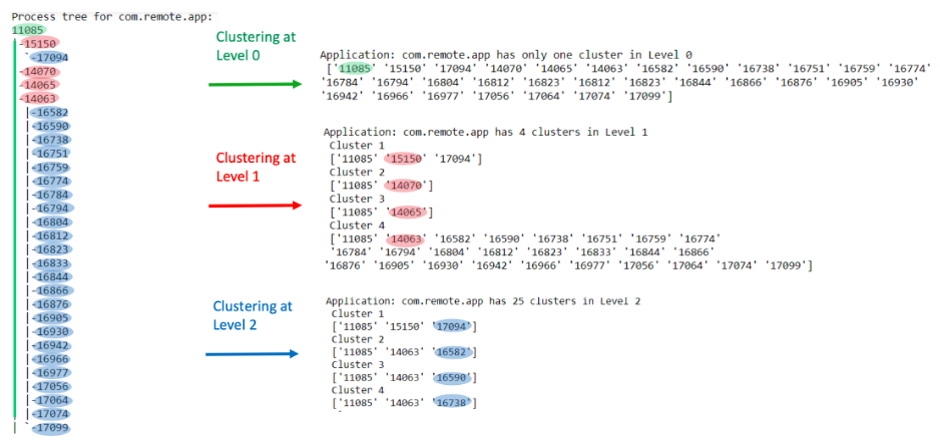}{Clustering of system calls. \label{clustering}}

The mobile ex-filtration dataset of interest is composed by features represented in the
form of 3-grams of system calls as seen in Figure \ref{preprocessing}. 

\Figure[!h][width=0.8\textwidth]{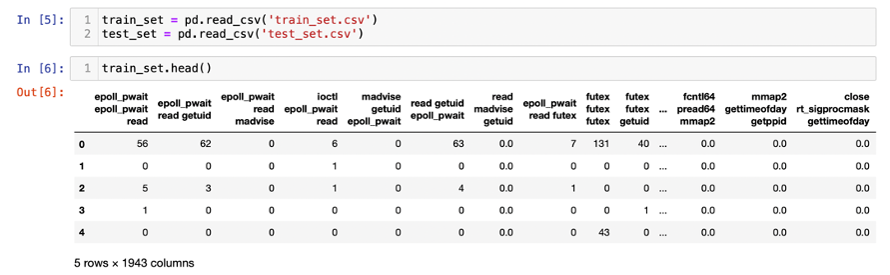}{Features as 3-grams of system calls. \label{preprocessing}}

\subsection{Discussion}
It is important to highlight the attacker’s capability in other scenarios and the assumptions imposed by the attacker, being sometimes to optimistic while in other scenarios represent more realistic conditions [72]. In the literature there are many examples related to assumptions in regards to the capability of the attacker. For instance, TRIM [76] assumes the ratio of the poisoning examples declared by the attacker as known. Deep-kNN [79] assumes access to ground-truth labels allowing the system to compare each sample’s k neighbors with the class labels.
\\
It is important to make a clear distinction on the properties of the crafted poisoning data, poisonous data obtained by performing label-flipping is one. Albeit, the scheme promises high accuracy degradation, it is far from representing the most effective option for an attacker. This is because label-flipping is considered among the most basic poisoning techniques and most of the existing defense mechanism can detect these ones as outliers and reject them with relatively ease.

\section{Methodology}
The dataset to be employed is the mobile exfiltration data in the work of Morcos et al. \cite{Martina}. This dataset is considered the main data of interest in this work as it represents the data used to feed varios  ML models. 
\\

The distribution of the data has been analyzed using the software Weka, having distinguished labeled samples in a histogram for each feature of the dataset. The results of this analysis show no features with separable data as it can be seen in Figure \ref{weka_separable}.

\Figure[!h][width=0.8\textwidth]{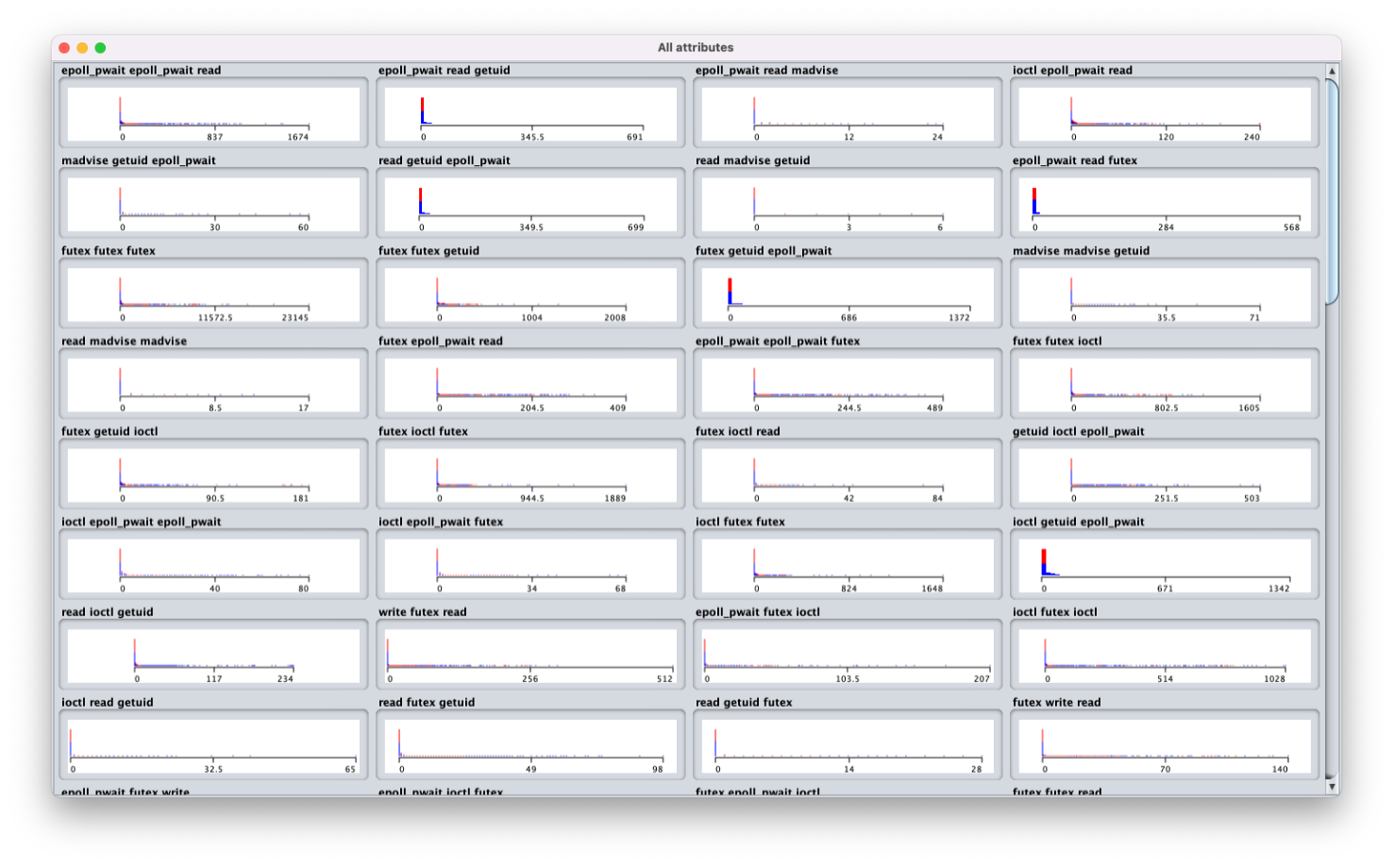}{Distribution per feature in malware dataset. \label{weka_separable}}

\subsection{Training ML classifiers}
Five different ML model have been selected as the target of the intended attack: Random Forest, Decision Tree, SVM, Logistic Regression and KNN. The development of these five ML classifiers for malware detection has been performed as case study, every ML model has been trained and tested with a part of the same mobile ex-filtration dataset. The dataset has been partitioned in 60\% for training, and 20\% for testing. The code can be found in Appendix A.1.
\\

The main purposes of this case study is to perform a comparison in performance, for this step we assume the existence of only clean data (ground-truth) with no existence of poisonous samples. The metrics of interest as part of the testing results are: Accuracy, Precision, Recall, F1 score and AUC (Area Under the Curve), the importance of the latter will be explained in the next subsections. The results obtained with each ML algorithm are shown in Figure \ref{clean_data_perfomance}. Being accuracy the most important metric for our study. Note that Random Forest is the model with the highest accuracy of them all, with an accuracy of 99.54\%, however it is important to remark that all the other machine learning models perform with an accuracy over 97\%, proving a high level of reliability among all the ML algorithms of interest.

\Figure[!h][width=0.45\textwidth]{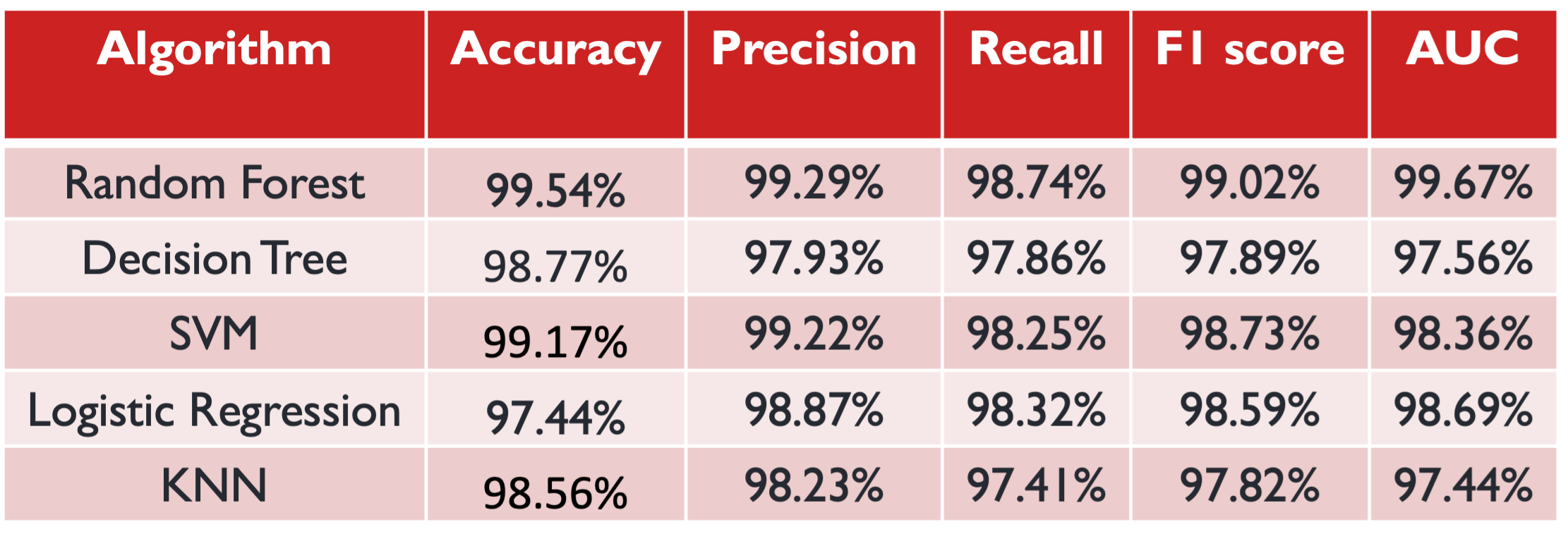}{Performance metrics of ML models. \label{clean_data_perfomance}}

The confusion matrix for each ML model is computed, this allows the correct visualization of the performance of every algorithm, comparing in 2 dimensions the number of actual and predicted results; true positives and true negatives, false positives and false negatives, respectively. It can be appreciated in Figure \ref{matrix} the number of miss-classified samples (false positives and false negatives) compared to the cases where predicted samples have been correctly classified according to their respective class.

\Figure[!h][width=0.4\textwidth]{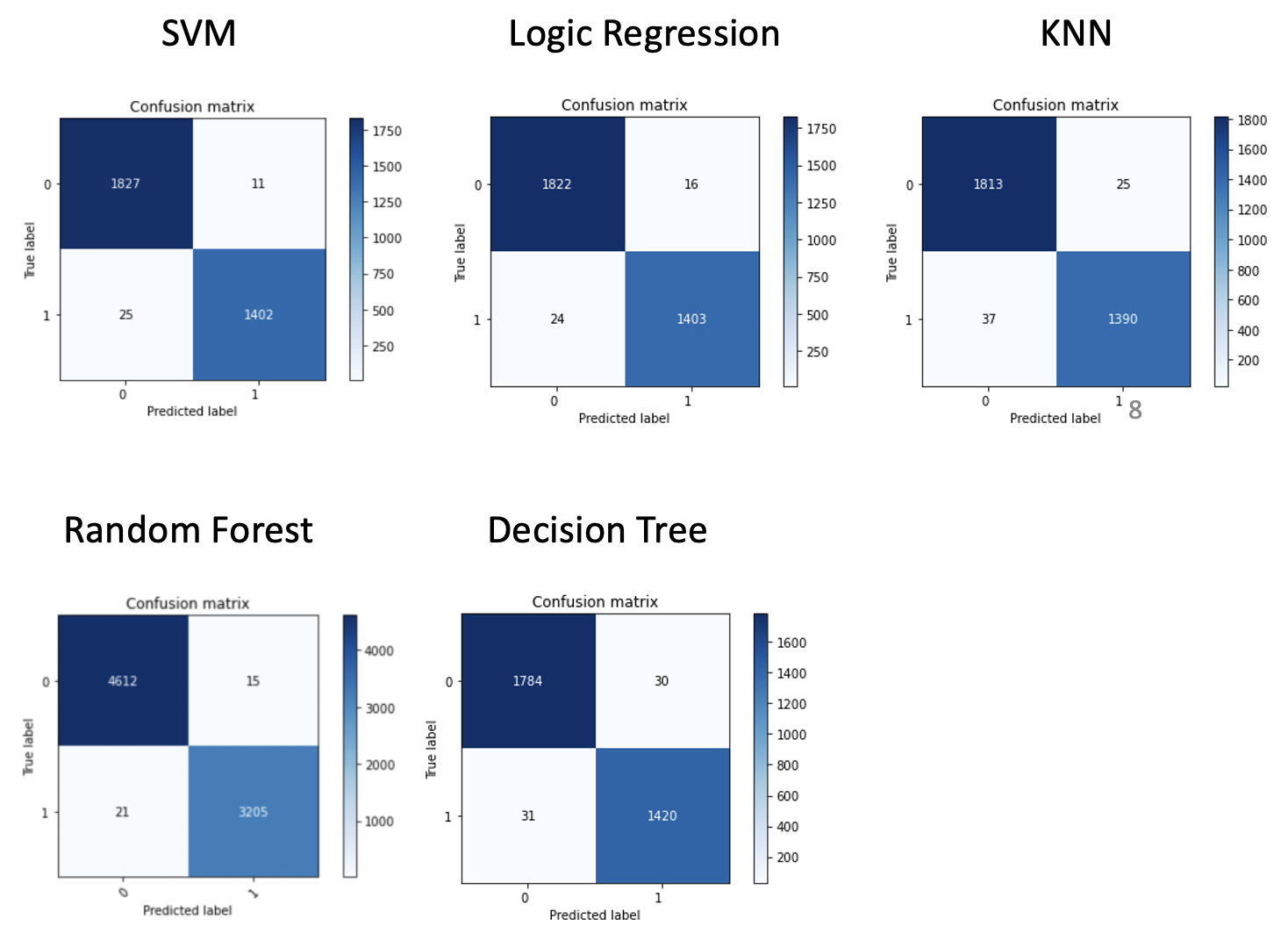}{Confusion Matrix comparison. \label{matrix}}

\subsection{Proposed data poisoning attacks}
The proposed attack approach is based on Label Flipping \cite{A02}, as explained in the previous chapter, this method tampers the labels of the samples injected into the training set, generating the poisoning data. The proportion of poisoning samples with respect to the total number of samples contained in the training set figures as an important parameter to evaluate, naturally the most efficient LF attacks are the ones that require the less number of poisonous samples, this has to do with the attacker capability to inject poison samples in high numbers.
\\

It is important to remark that the target ML models of interest have been trained in the previous steps with clean data, assuming no poisoning samples. Then, the proposed attacks explained in the following sections will be acting during the re-training of the ML model, then a comparison between its original performance (without poisoning samples) and the performance of the model after the attack in question can be further analyzed.

\subsubsection{Label-flipping attack: Level 1}
A random label-flipping attack is one if not the simplest type of attack to craft against a ML classifier, nonetheless it is considered one of the attacks that can cause the most damage to mostly any ML classifier’s accuracy. It consists in switching the labels associated to each sample in a random way and inject them into the training set of the model. The mobile ex-filtration dataset contains a feature indicating the label of the sample, this can be either Benign “0” or Malign “1”.
\\
The chart in Figure \ref{acc_compare} and \ref{table_comparison} well serves as a comparison between the model original performance (0\% poisoning samples) and the performance of the model after the attack in question can be further analyzed, the results are displayed by performing a variation in the proportion of poisoned samples, accounting for a 25\%, 50\%, 75\%, 100\% poisoned samples.

\Figure[!h][width=0.8\textwidth]{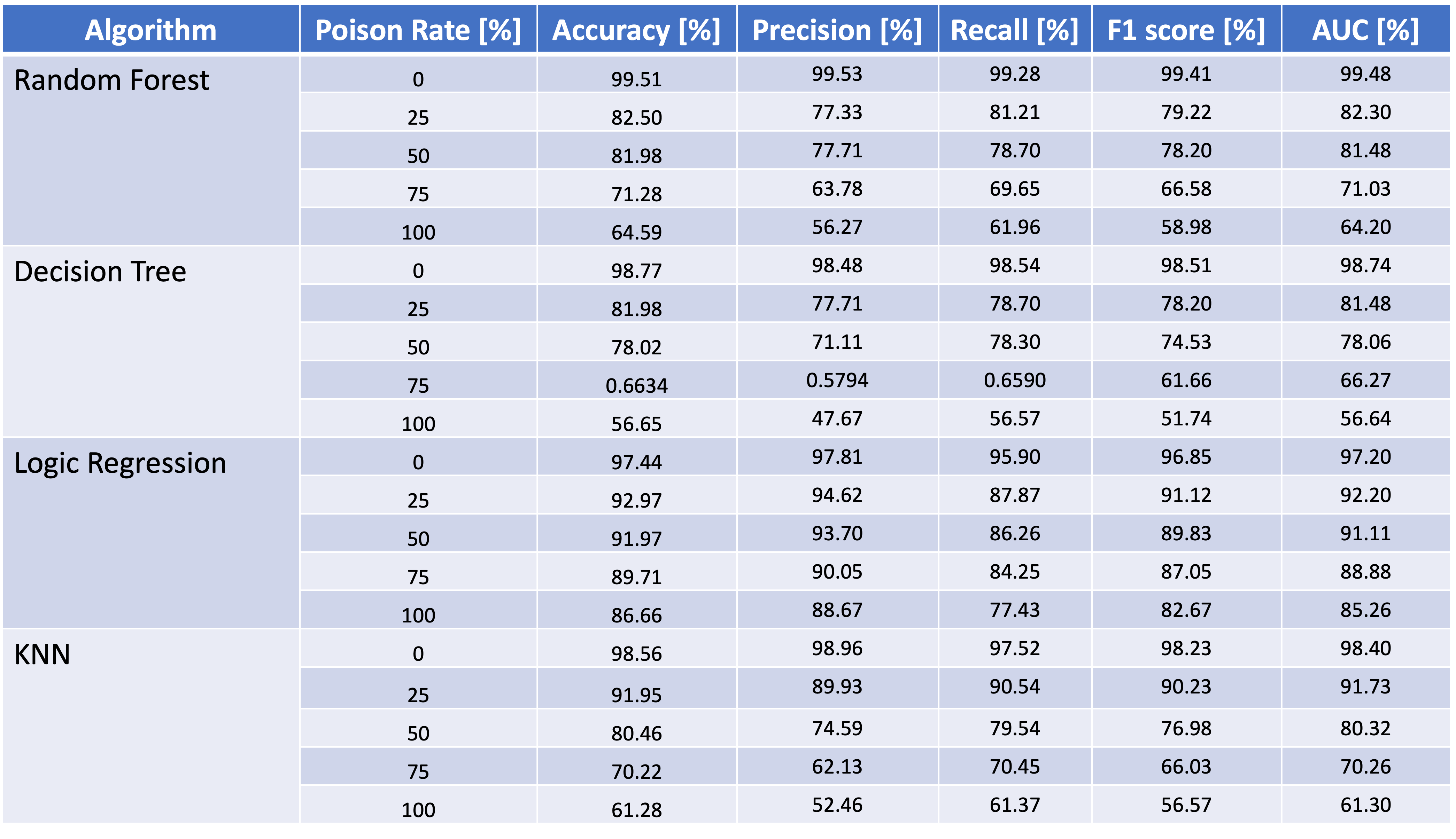}{Metrics comparison among ML models \label{table_comparison}}

\Figure[!h][width=0.45\textwidth]{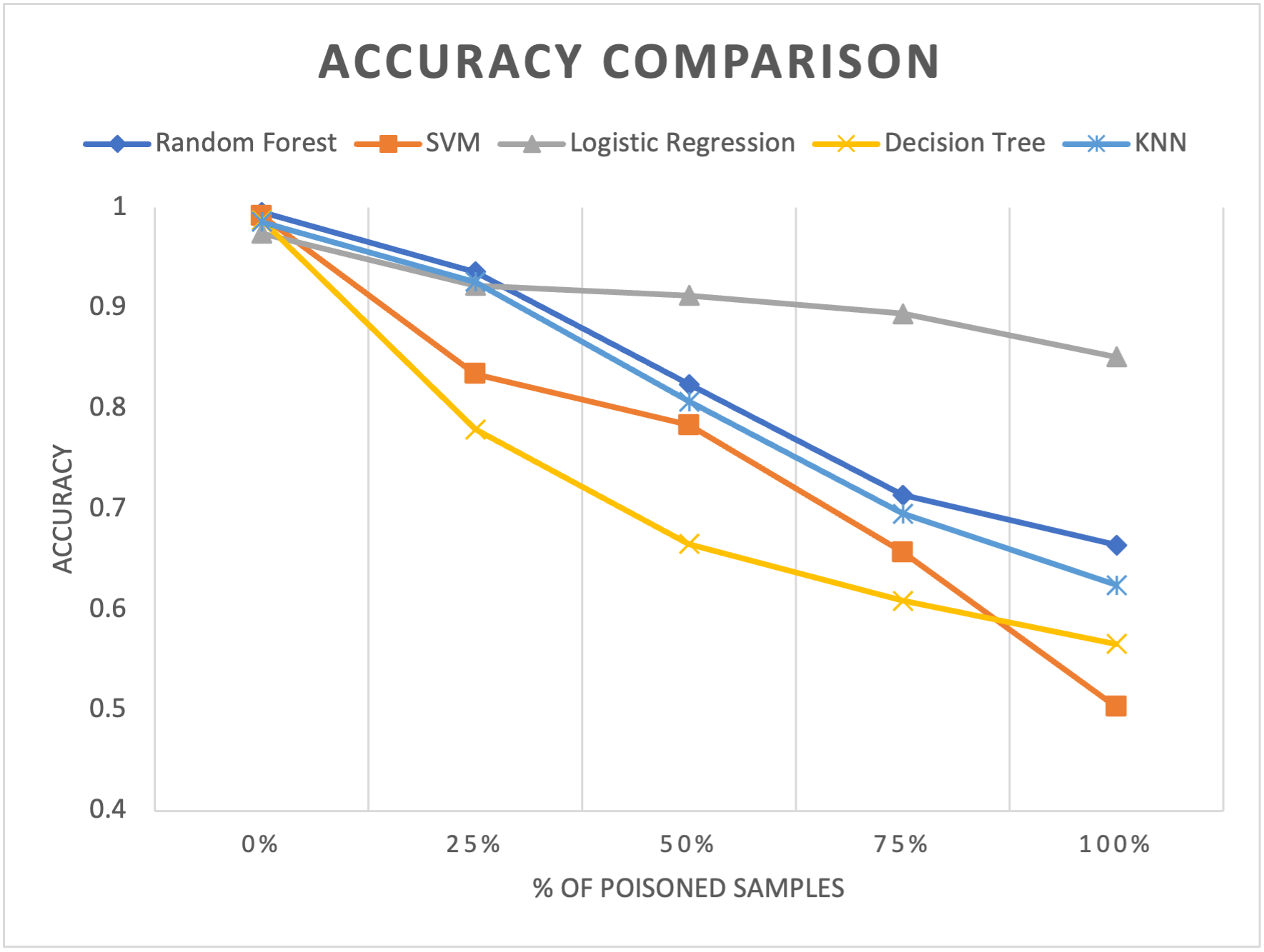}{Accuracy drop comparison among ML models \label{acc_compare}}

The series of steps necessary to perform the proposed attack are explained in better detail in the form of a process map in Figure \ref{process_lf_1}.
\\

\Figure[!h][width=0.7\textwidth]{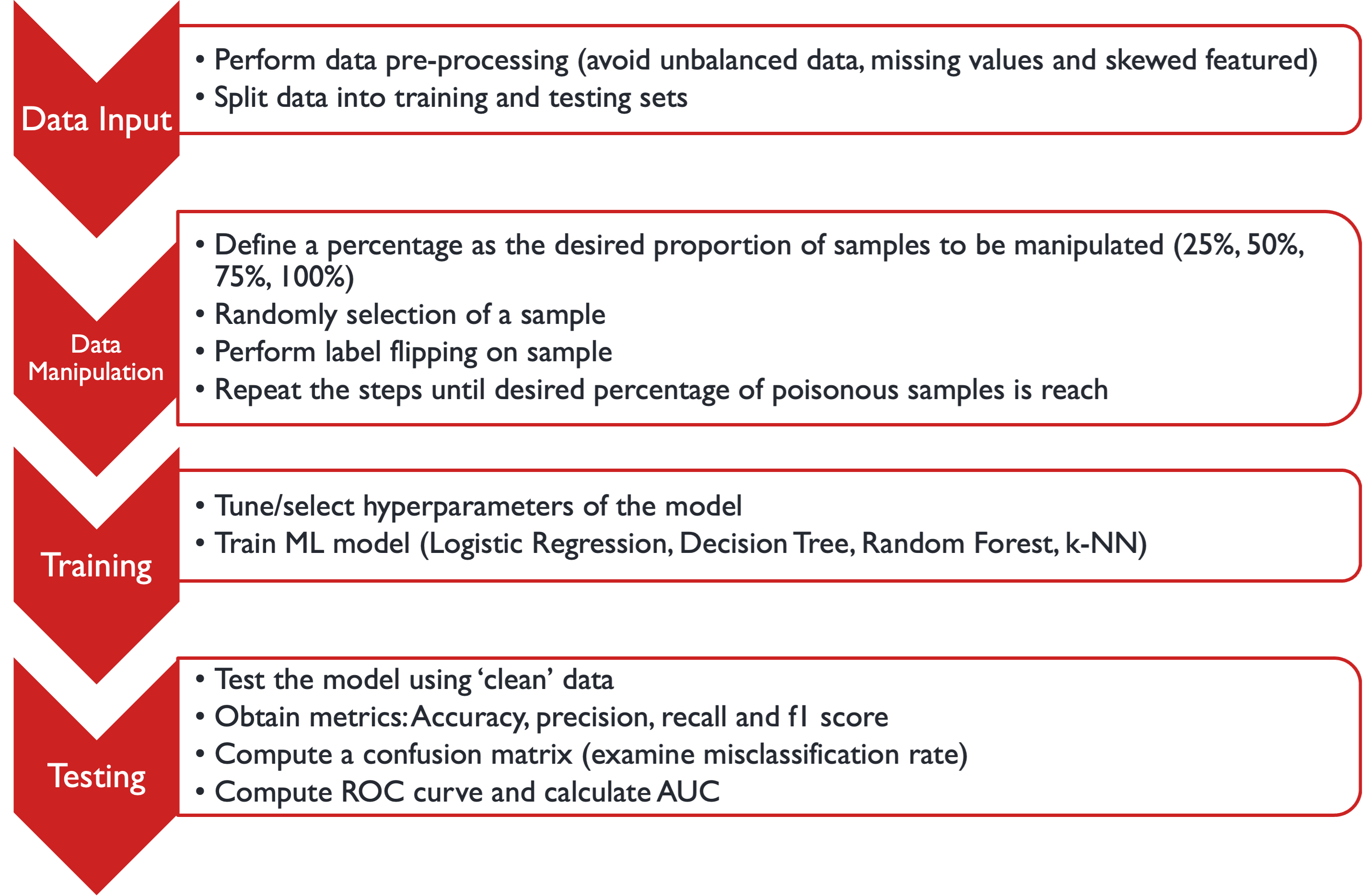}{Process map: DP attack: Level 1 (Random Label-flipping) \label{process_lf_1}}

We have computed the  confusion matrix for each ML model. For this case we are interested in reporting the results of the confusion matrix when accounting with 'clean' data only and also under the each poisoning scenario when varying the ratio of poisoning samples. The intention with this is to appreciate and compare clearly the steady increase in the miss classification rate by the model when being presented an increased number of poisoning samples. The effects of the overall missclassification of the two classes of interest have been studies and reported for each algorithm, comparing the effects of the attack with the 'clean data' condition (no poisoned samples present).For this please refer to Figures \ref{dt_matrix}, \ref{rf_matrix}, \ref{lr_matrix} and \ref{knn_matrix}.

\Figure[!h][width=0.4\textwidth]{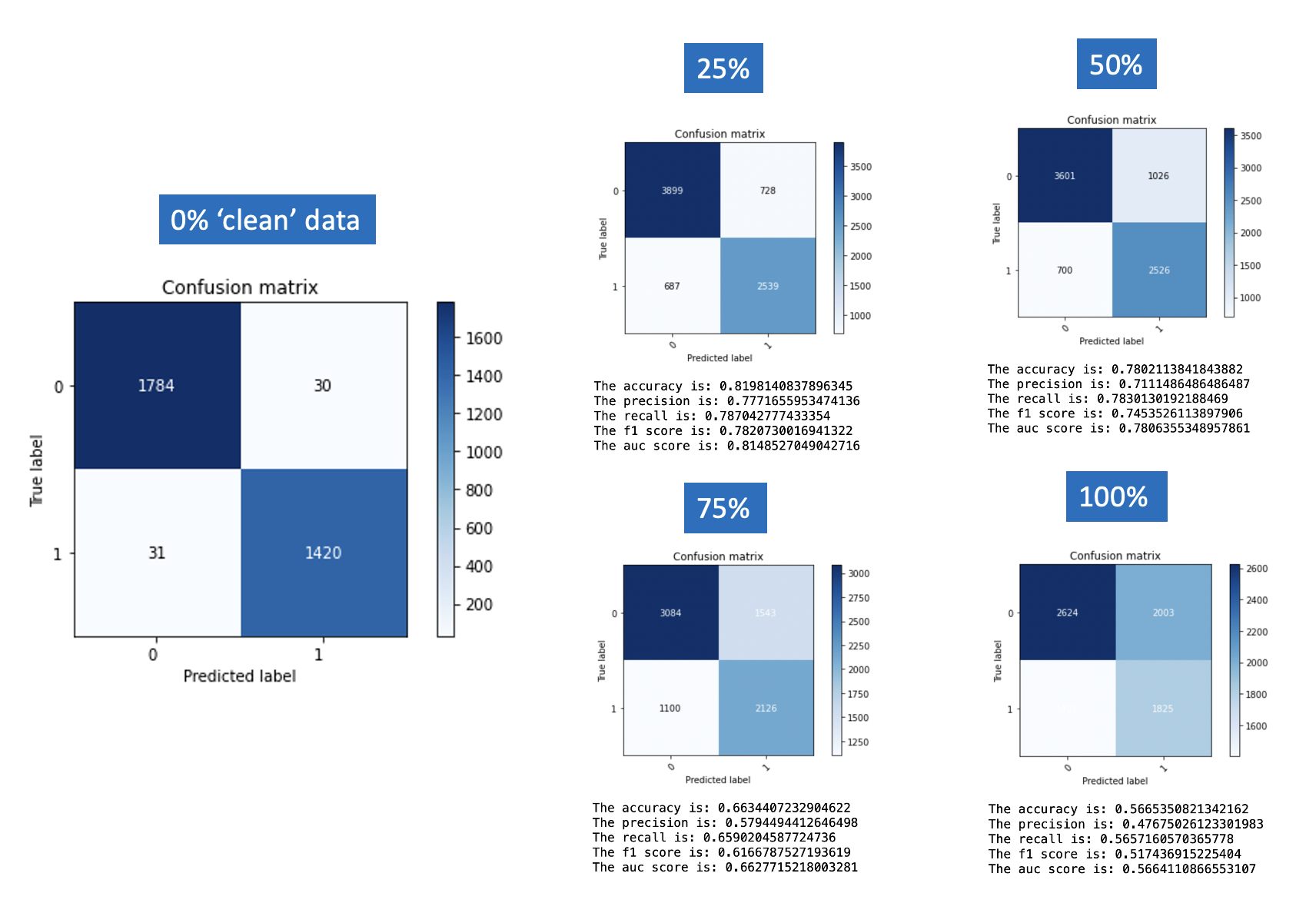}{Decision tree confusion matrix comparison \label{dt_matrix}}

\Figure[!h][width=0.4\textwidth]{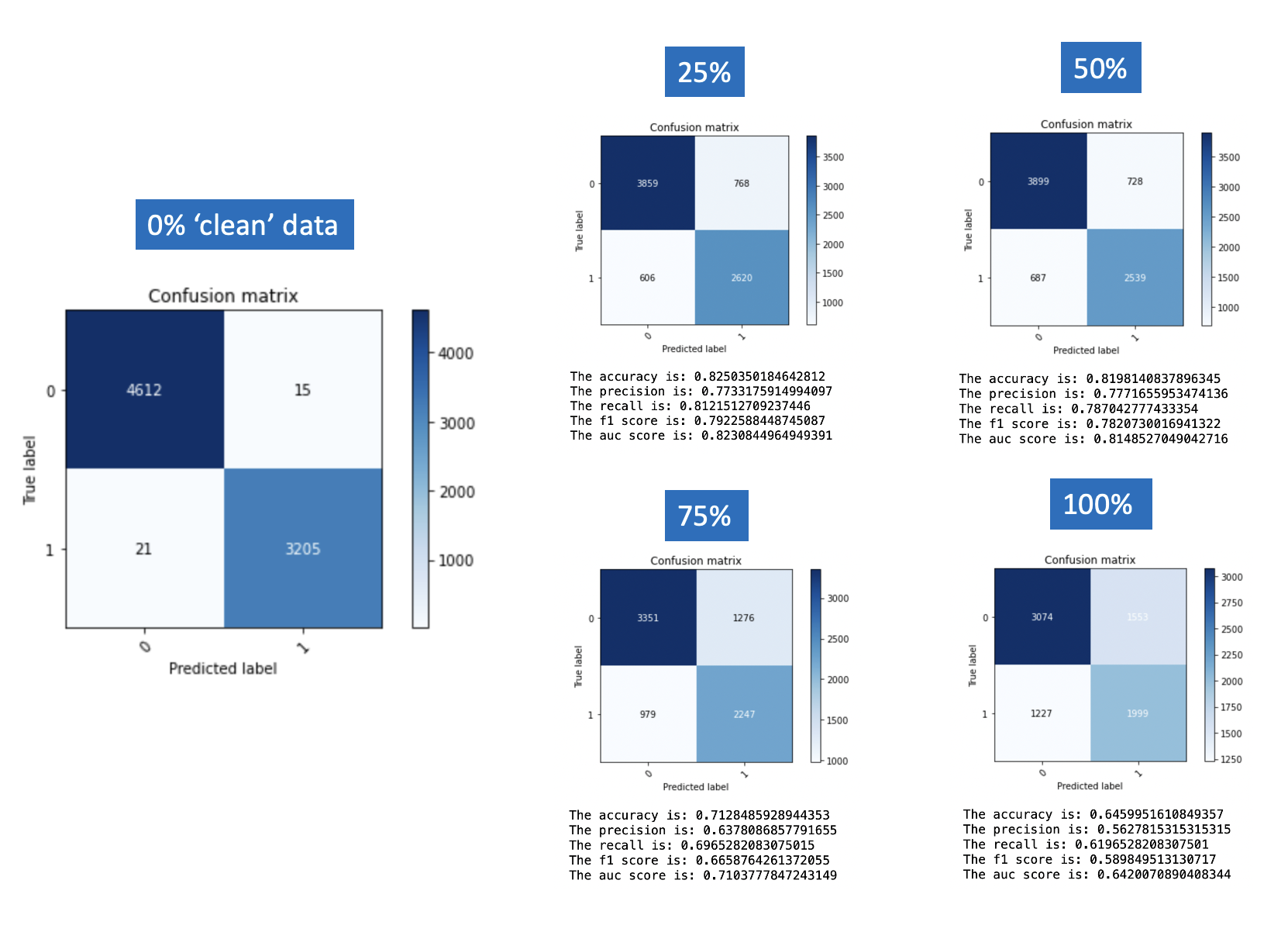}{Random forest confusion matrix comparison \label{rf_matrix}}

\Figure[!h][width=0.4\textwidth]{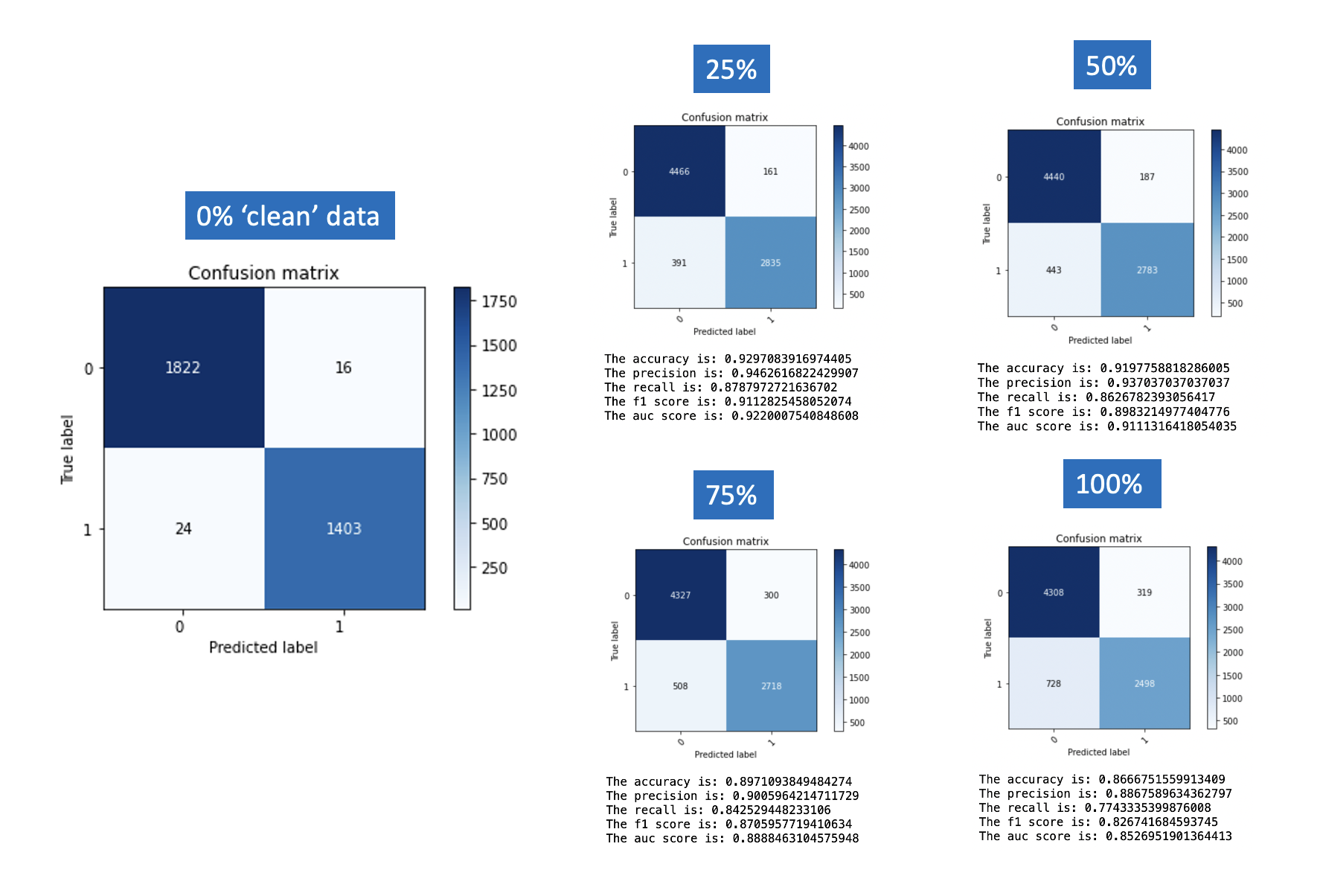}{Logistic regression confusion matrix comparison \label{lr_matrix}}

\Figure[!h][width=0.4\textwidth]{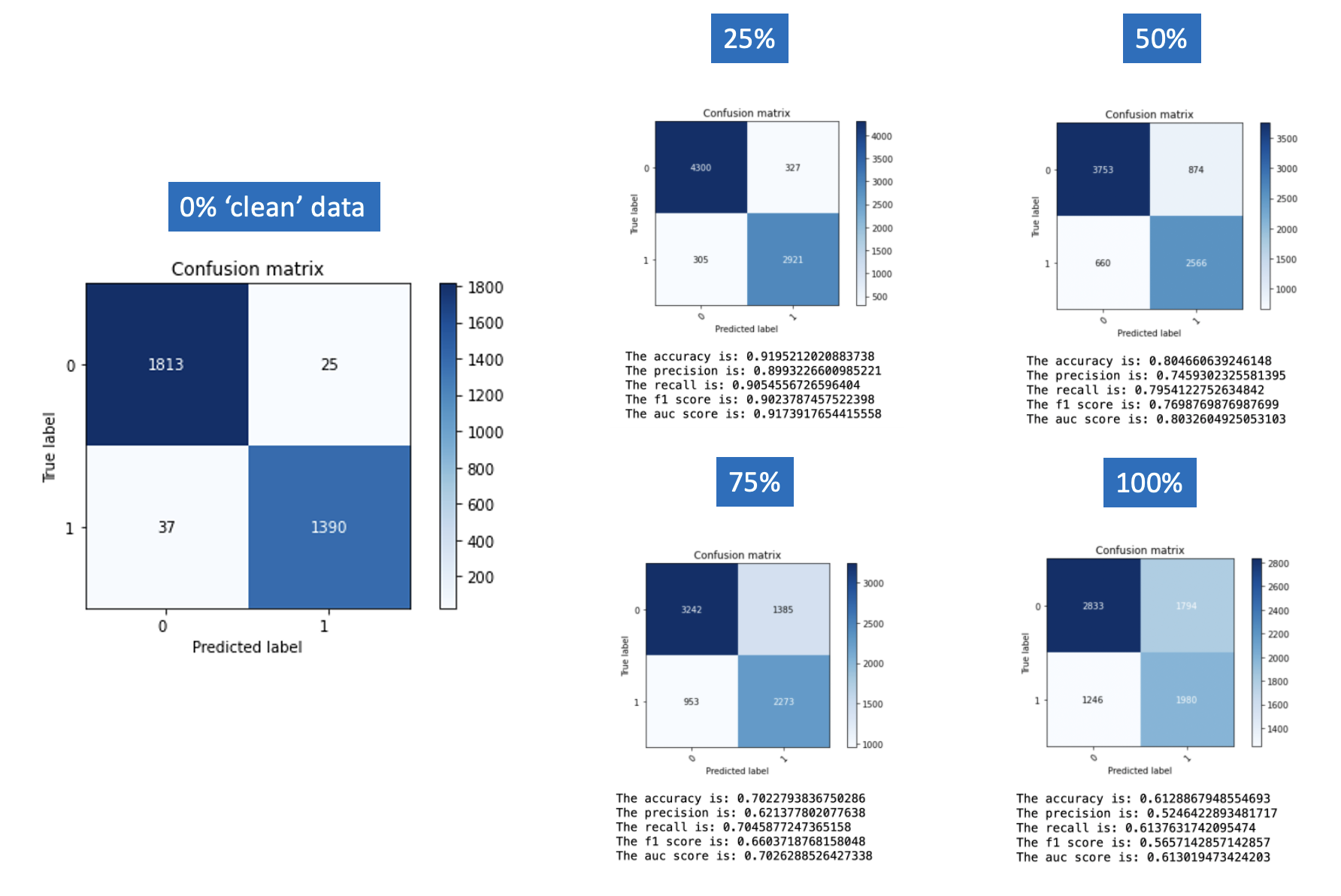}{KNN confusion matrix comparison \label{knn_matrix}}

The same comparison can be made when computing the ROC curve for each model under 'clean' data assumption as well as for the same poisoning scenarios already described before. Then a degradation in the AUC by increasing the ratio of poisoning samples is visible, indicating a reduced true positive rate with an increased false positive rate. This behaviour can be better appreciated in Figures \ref{dt_roc}, \ref{rf_roc} and \ref{lr_roc}.

\Figure[!h][width=0.45\textwidth]{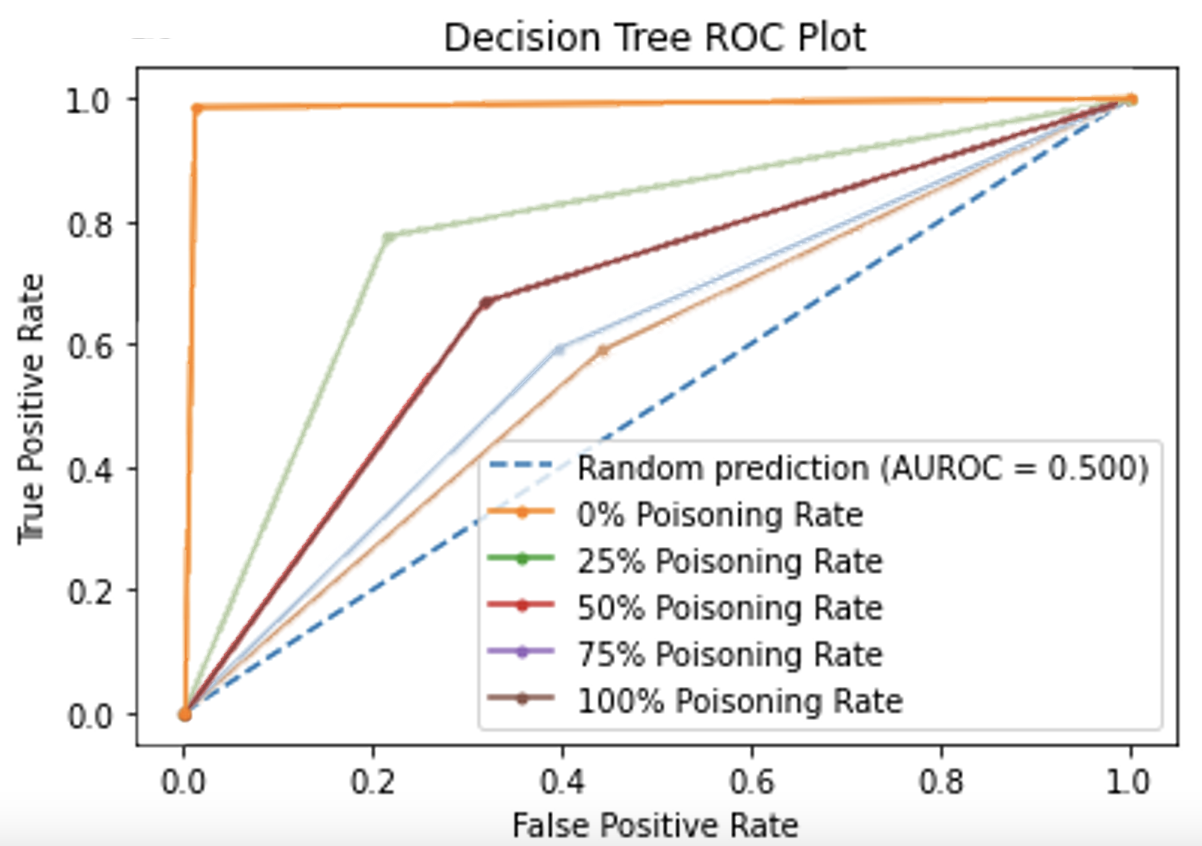}{Decision tree ROC curve comparison \label{dt_roc}}
\Figure[!h][width=0.45\textwidth]{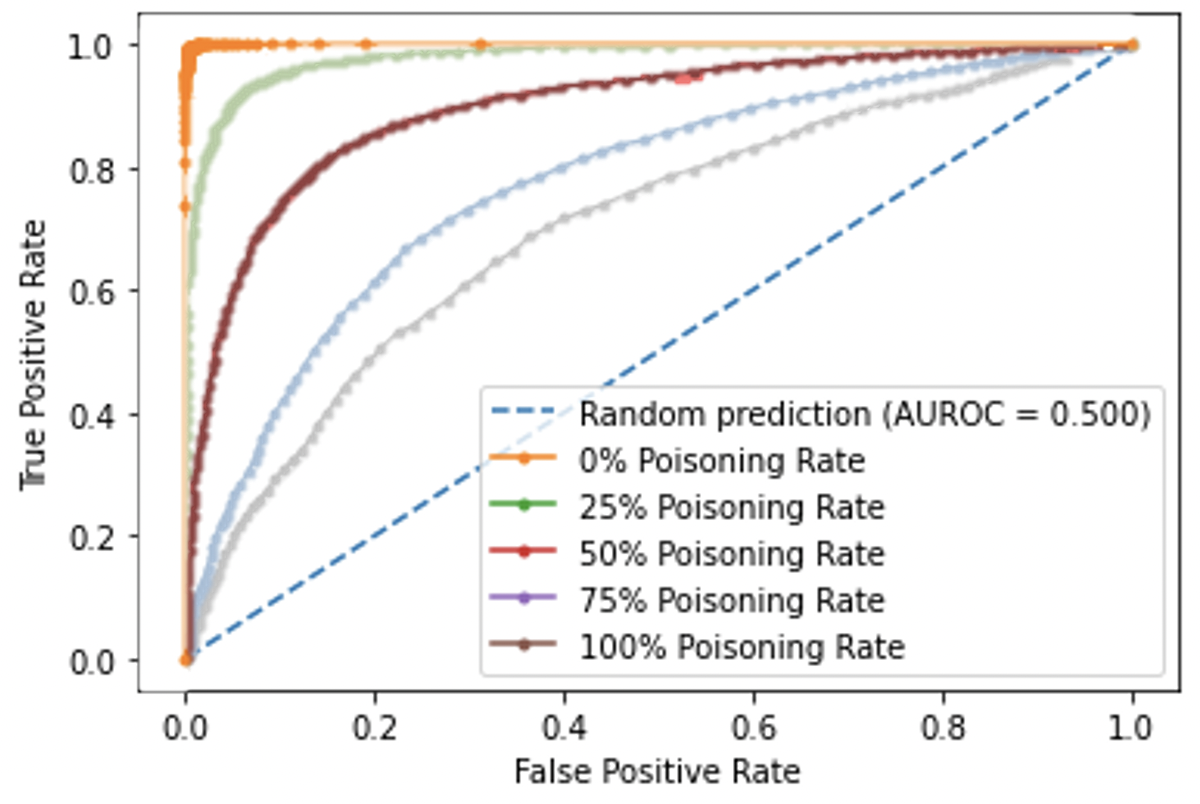}{Random forest ROC curve comparison \label{rf_roc}}
\Figure[!h][width=0.45\textwidth]{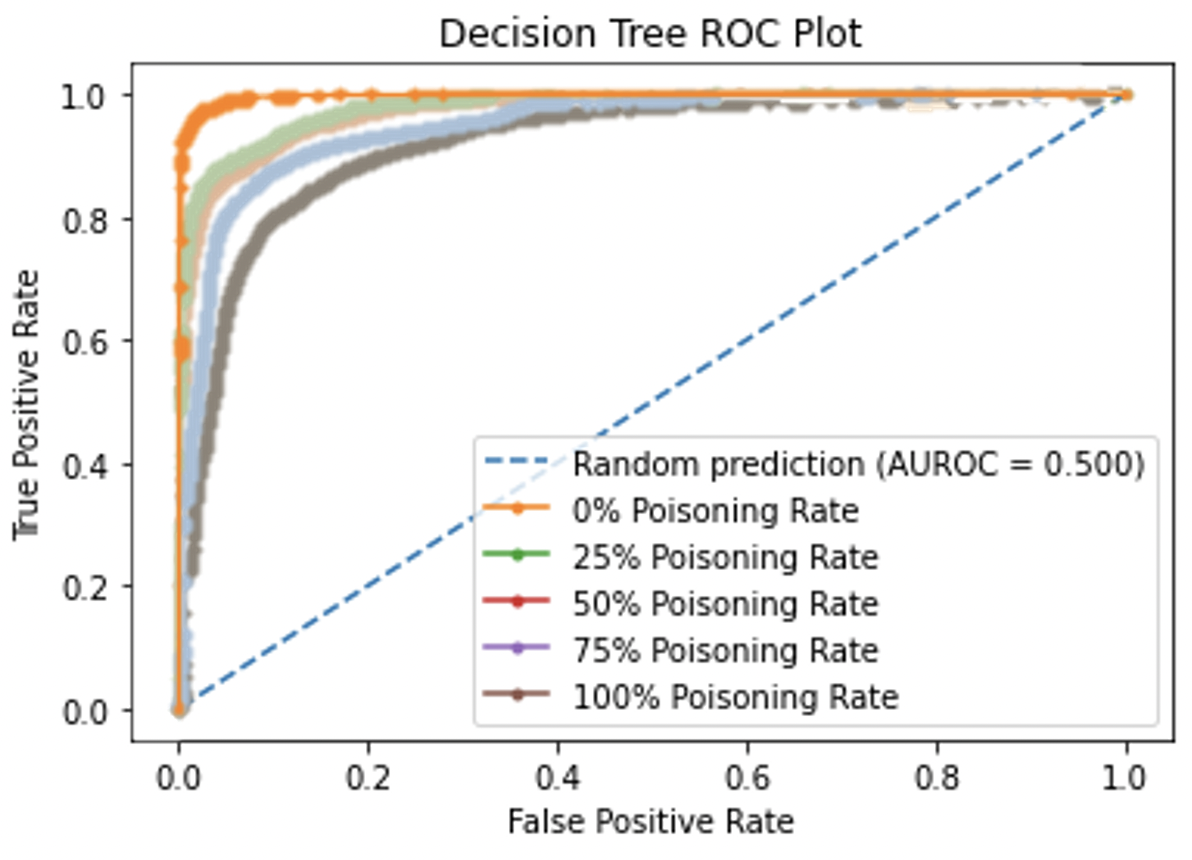}{Logistic regression ROC curve comparison \label{lr_roc}}

\subsubsection{Label-flipping attack: Level 2}
As a second line of attack, rather than randomly, the aim now is to perform label flipping targeting a specific class \cite{A02}. Then causing a significant reduction in the accuracy of the target model is no longer the priority of the attacker, but the primary aim of the attack is to deliberately achieve the misclassification of a determined class in particular. Naturally a potential attacker will opt to misclassify the malign malware samples by switching their labels from 'malign' to 'benign'. The aim of the attacker in question is to tamper the model’s ability to recognize and classify malign malware as ‘malign’, thus letting through any type of threatening malware to a computer system. 
\\
In consequence, during the testing phase the machine learning model will misclassify malign samples as benign instead, tampering completely the decision-making process of the classifier in question. In this case, the assumption of the attacker is the following. The attacker must have prior access to the dataset and know the set of the features in the dataset. Then there's is the possibility for the attacker to employ Explainable AI in order to determine the feature/predictor importance among all the ones in the dataset and figure out the number of features that generate the most impact in the classification task, this classification task is directed towards the samples that are labeled as 'malign'.
\\
\Figure[!h][width=0.8\textwidth]{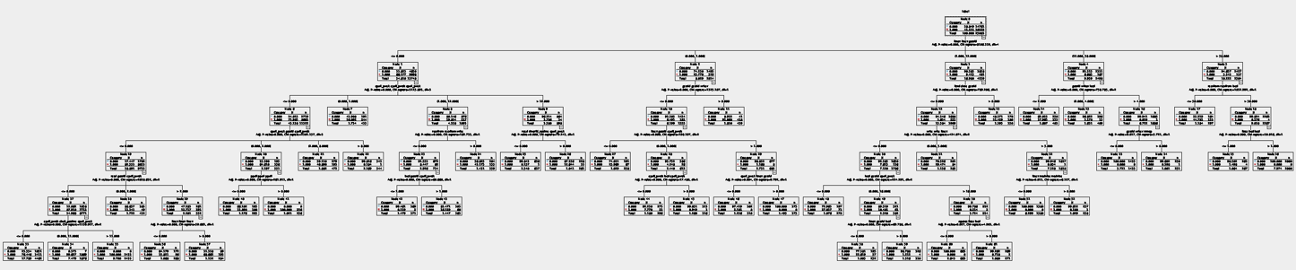}{Explainable AI using generated decision trees \label{generated_trees}}

For this purposes, we have employed Explainable AI to emulate the process a potential attacker will undergo by generating decision trees (Figure \ref{generated_trees}) using the tool IBM SPSS Modeler and identify the features with the highest importance for the classifier. This features can be seen in the Pareto diagram in Figure \ref{pareto_features}.

\Figure[!h][width=0.45\textwidth]{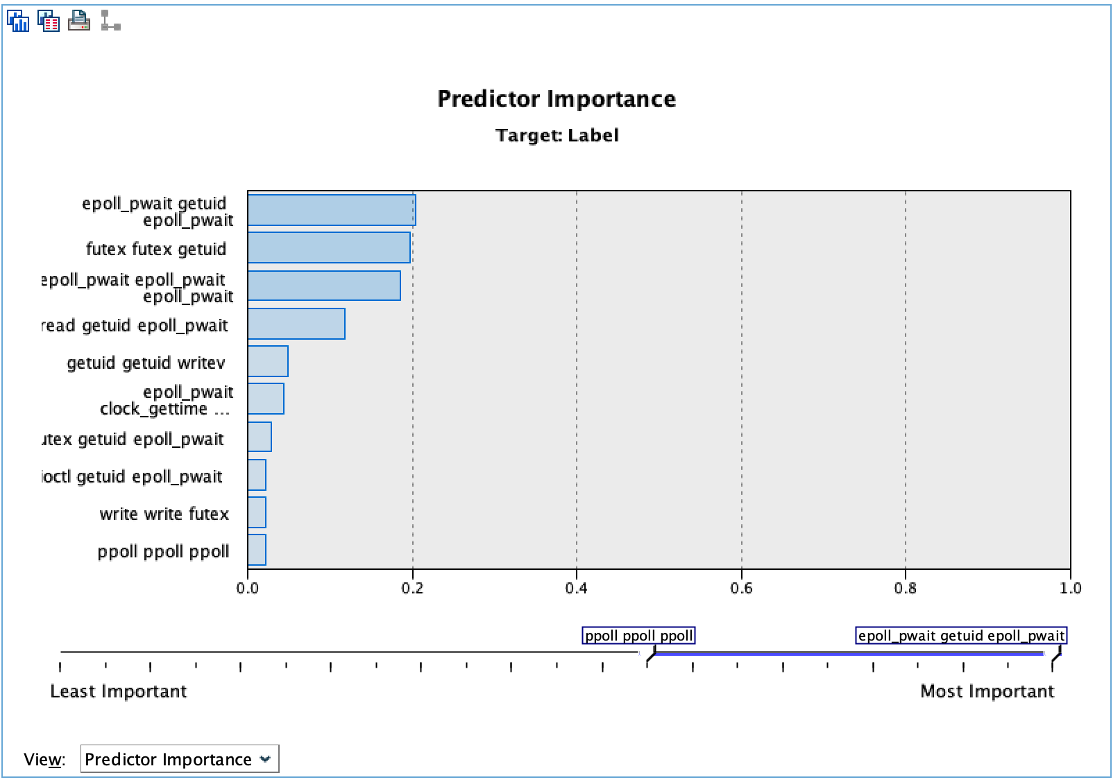}{Pareto diagram: Predictor/feature importance \label{pareto_features}}

We will refer to the process map shown in Figure \ref{process_map_lv2} to better understand the algorithm. Once selected the features of interest we have defined a 'decision criteria' to determine the samples subject to the attack. This decision criteria is based on the following conditions.

\begin{itemize}
\item Sample must be originally labeled as 'malign'.
\item The sample must have values associated to the determined features of interest (monogram or 3-gram of system calls) that overpass a determined threshold. In this case we have defined that threshold as any value that is more than "0".
\end{itemize}
\hfill

\Figure[!h][width=0.7\textwidth]{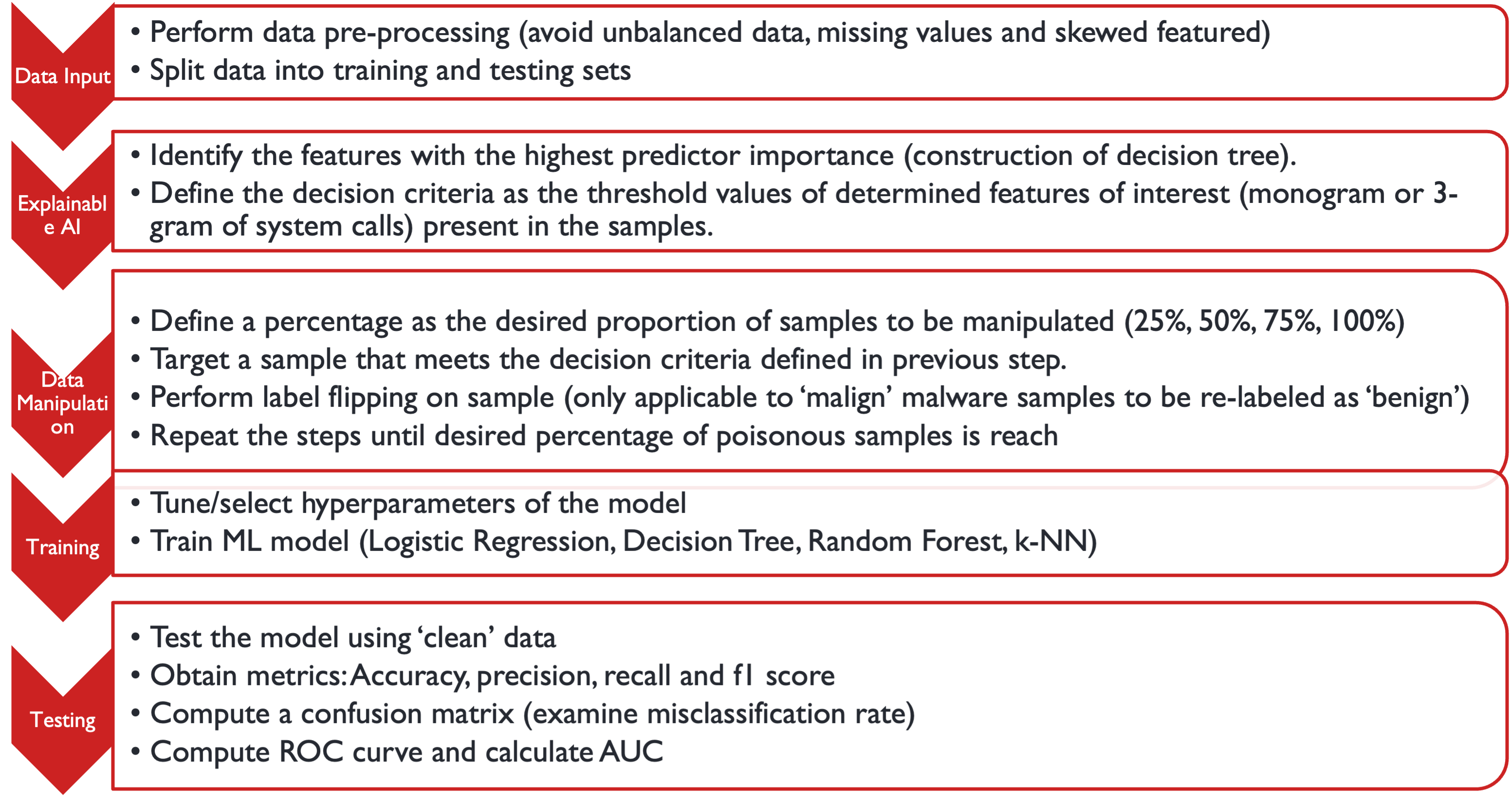}{Process map: DP attack: Level 2 (Target label-flipping) \label{process_map_lv2}}

The chart in Figure \ref{attack_level_1_level2} well serves as a comparison between the model original performance (0\% poisoning samples) and the performance of the model after the proposed attack. As we did for the label flipping attack: Level 1. Again, the results are displayed by performing a variation in the proportion of poisoned samples, accounting for a 25\%, 50\%, 75\%, 100\% poisoned samples. For this case Figure \ref{attack_level_1_level2} includes the results of the DP attack: Level 1 as a mean of comparison.

\Figure[!h][width=0.45\textwidth]{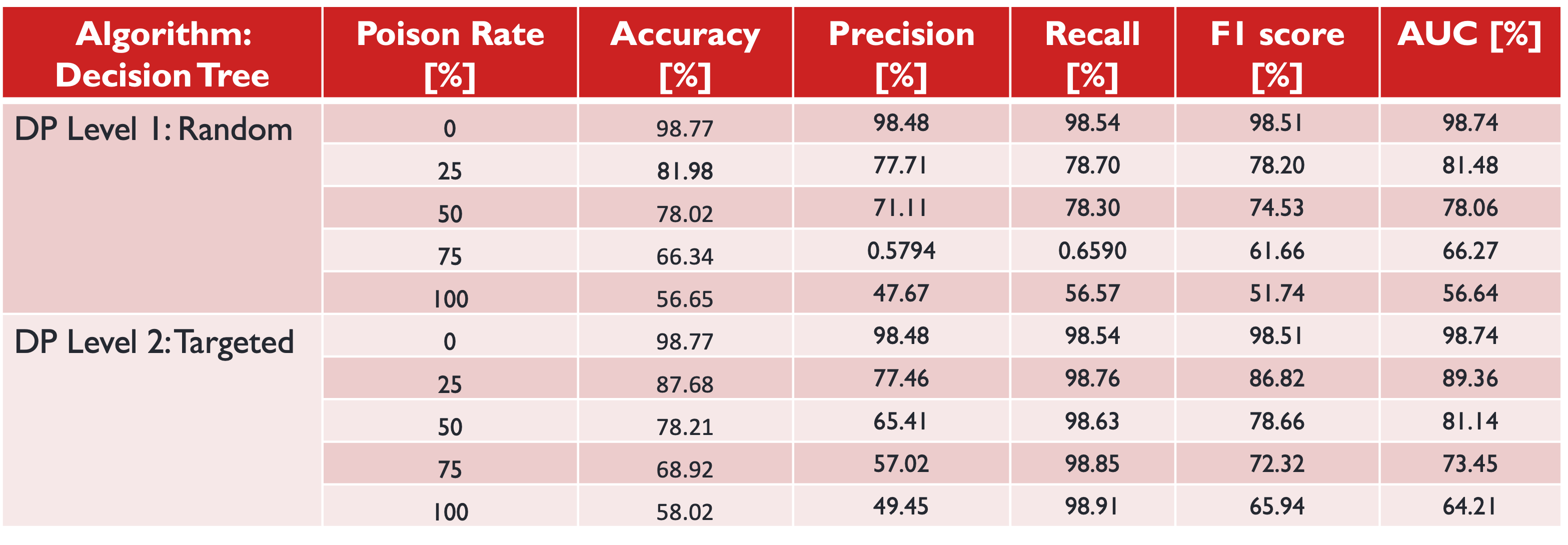}{Metrics comparison among ML models after DP attack: Level 1 and DP attack: Level 2 \label{attack_level_1_level2}}

Notice that the effect of such an attack will not only affect the accuracy of the model but also to increase dramatically the number of false negatives compared to false positives, this behavior can be better appreciated in the confusion matrix for the decision tree algorithm in Figure \ref{dt_cm_lv2}. 

\Figure[!h][width=0.45\textwidth]{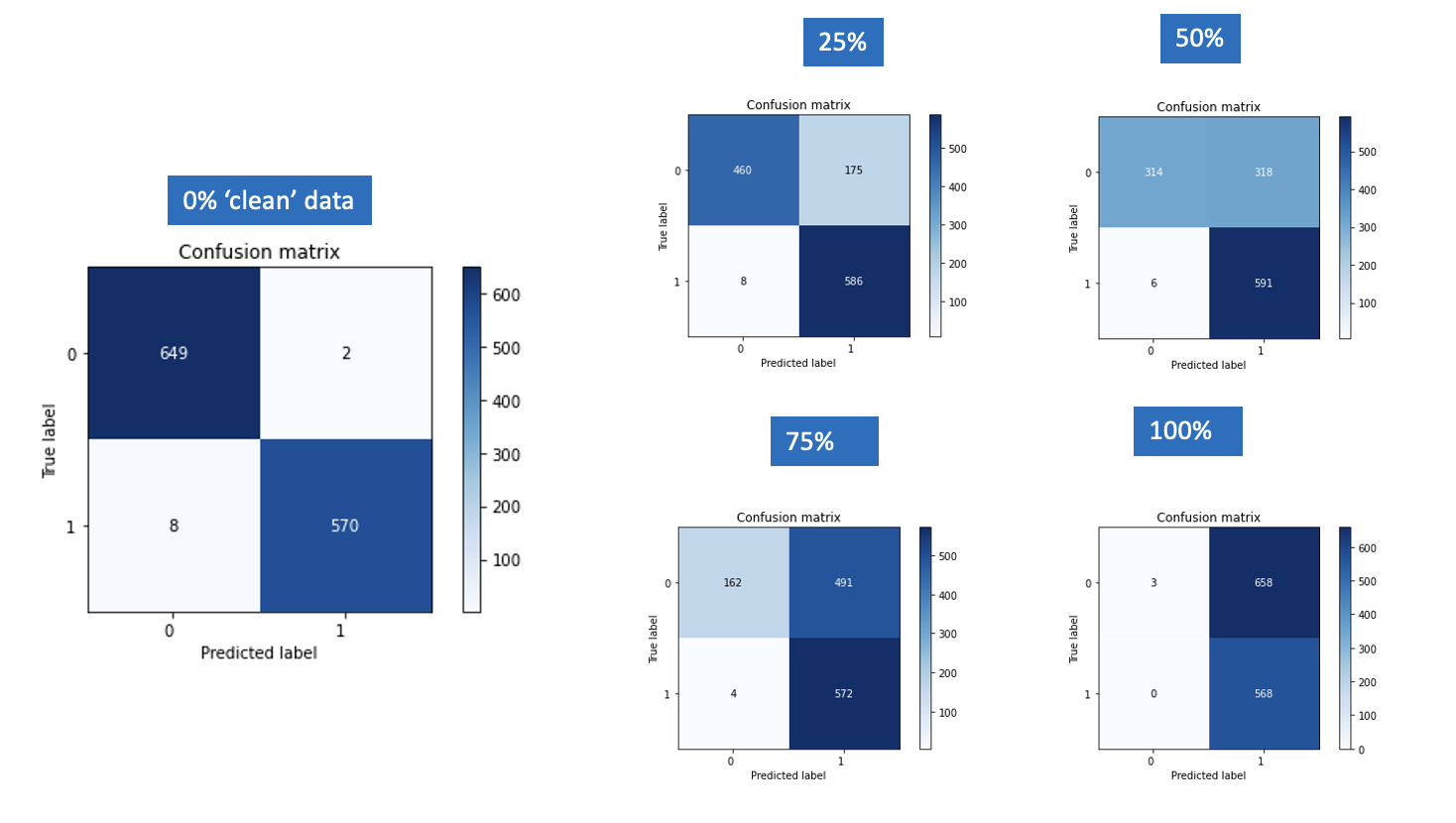}{Decision tree confusion matrix comparison under Level 2 DP attack (Target label-flipping) \label{dt_cm_lv2}}

This impacts specially the recall metric which is associated to the number of false negatives, as in the previous attack (Level 1: random label-flipping) this can be better appreciated by comparing the generated ROC curves in Figure \ref{dt_roc_lv2}, this ROC depicts the perfomance of the decision tree algorithm.

\Figure[!h][width=0.45\textwidth]{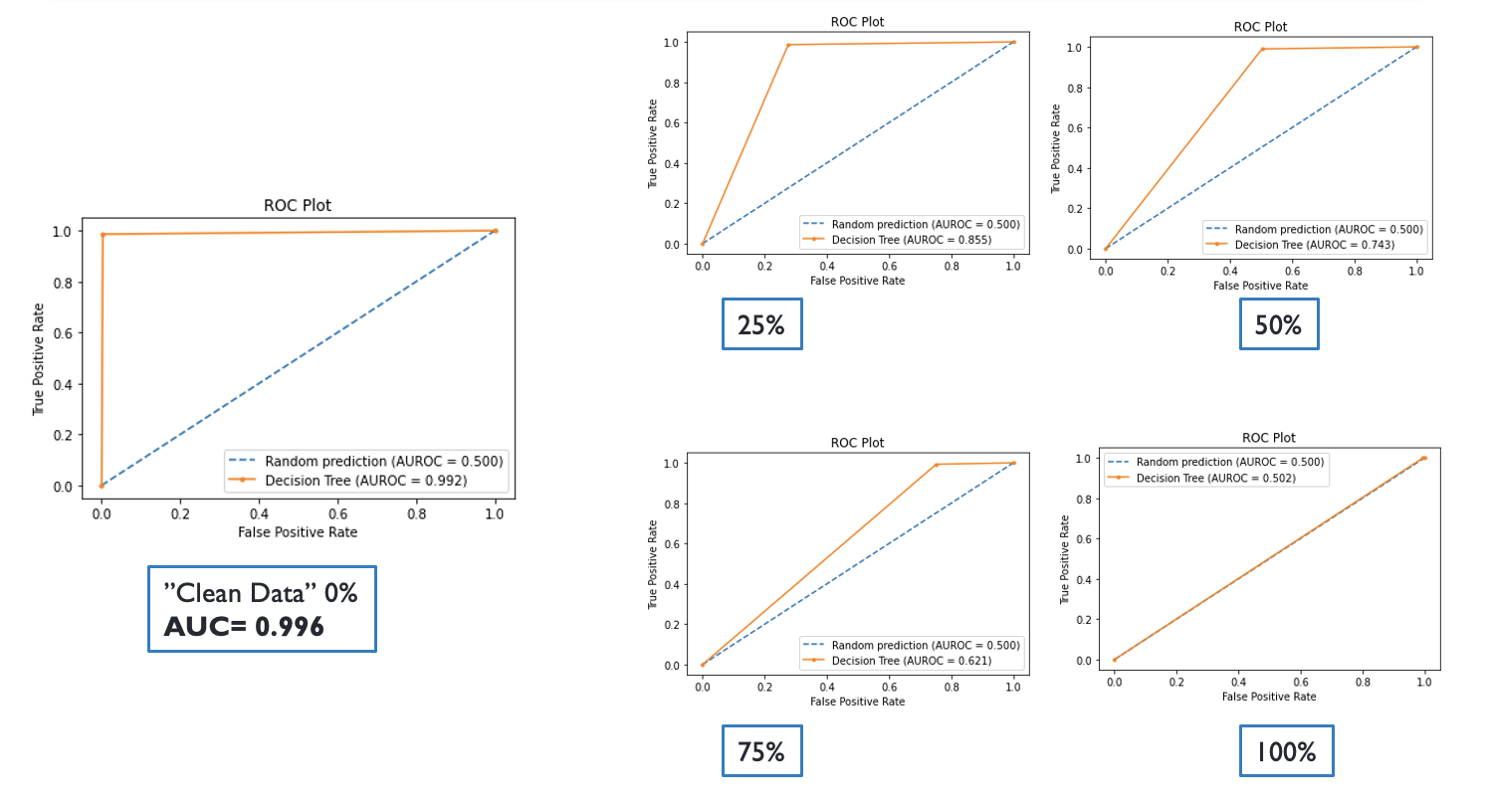}{Decision ROC curve comparison under Level 2 DP attack (Target label-flipping) \label{dt_roc_lv2}}


\subsection{Discussion}

Nowadays, there is an urgent need of an attack-agnostic defense system, since most of the works in the literature contemplate combating one type of attack in specific. Nonetheless label flipping still represents an important way for benchmark to assess the robustness of any given type of ML classifier. Therefore, all the proposed approaches have been studied theoretically as viable candidates for a defense against DP. 
\\

Label flipping attacks nowadays still impact negatively machine learning classifiers once identified new vulnerabilities in specific scenarios related to specific applications, the nature of the dataset of the malware detector that is being studied could promise an outstanding area of opportunity. In addition, label-flipping still represents an important way for benchmark to assess the robustness of any given type of ML classifier. 
\\

Therefore, the intended research proposal aims to craft an attack more than just one type of machine learning model and analyze their response and drop in performance. Particularities regarding the decision-making process of each algorithm will come into play, nurturing posterior analysis and further insights around the features involved in the malware dataset. Once having exploited and analyzed all upcoming vulnerabilities, an algorithm for a defense approach will become more clear and more suitable alternatives suited for either detection of mitigation tasks will be proposed, developed in further detail and tested in conjunction with different ML classifiers in order to assess their effectivity and feasibility. 
\\

\section{Future Work}
\label{sec:future work}
Nowadays ML models can no longer be seen as black box systems to be assessed solely on their results, but nowadays a deep understanding of the model is necessary in order to identify security flaws that, if not properly addressed, could lead to critical undesirable outcomes. Any breakthrough in the field of machine learning will always implicitly allow the introduction of new vulnerabilities. Such vulnerabilities will always pose an open window of opportunity for adversarial entities to exploit, leading as a consequence to an opportunity to develop new defenses counterbalance the outcome in the same matter as it occurred with the invention of the internet and the introduction of cyber-security systems.
\\

Assumptions for both proposals for an attack and defense will be imperative. This will imply reporting the specific details regarding the required knowledge over the ML model and the training set needed to attain a successful attack over a classifier. Similarly, it will become substantial to describe the specific instances in which the defense will be deployed. Regardless of the nature of our proposed defense system, it is important to maintain an open mindset; one which might well consider very optimistic or ideal scenarios, as well as others not so ideal, but rather practical for real life applications. In relation to this last point, we could add as an special consideration that; in contrast to other works, our work might not be suited to be assessed or tested with a common malware dataset, but rather with another kind of malware exfiltration engine, then a this work could promise great potential to become a novel solution.
\\

In this section we propose several potential and possible solutions to counteract the proposed series of label flipping data poison attacks depicted in this work. For this, we will showcase the different approaches that could be followed in order to propose a defense mechanism.

\subsection{First Approach: Detection and Mitigation}
As seen in random label flipping attack type 1, the accuracy drop did not overpass 50\%, this is due to the presence of low confidence points near the decision boundary, regardless of their assigned label (benign/malign). Such behavior could be the result of the nature of our binary classifier and needs further study in the short term to understand the ML decision process in better detail. Then this might be a good area of opportunity. As a first step, it could be possible to use the data gathered from the five ML models of interest once poisoning the data to find an acceptable threshold that could be used to reject this data and remove these points from the training set, discarding them as outliers. 
\\

Then, this approach will serve as a detection mechanism, being able to accept or reject the samples once identifying them within the acceptable threshold. Thereby detection techniques are meant to spot abnormalities in the form of noisy points (outliers) or due to the presence of poison samples themselves, this often requires constant monitoring of the target model observing the effects of the poisoning data over the performance on the model.
\\

On the other hand, there is another approach known as defense mitigation. A defense system capable of mitigating the effects of data poisoning, reducing the overall classification error as a result. A defense mitigation technique is very suitable since it is considered a proactive defense instead of a reactive one, being the first a technique able to generalize better, accounting for a wider variety of attack scenarios to suppress. Nonetheless it will be important to remark that this type of defense mechanism often entails a more sophisticated process than detection defense mechanisms. Albeit a mitigation defense mechanism may not exclude poisoned samples as part of the training set, it certainly involves a more sophisticated development process. This is because viable solutions for mitigation sometimes are more fundamented into probability of accounting for mislabeling samples than relying on the training labels alone. An example of this is seen in [70] wherein multi-objective optimization have represented a solution to this problem for SVM classifiers.

\subsection{Second Approach: Data Sanitization}
Similarly, to the previous idea. Another approach could seek detecting malicious data and counteracting the effects of the same before getting the sample points to the training set, such approach commonly receives the name of Label sanitization (LS) as seen in the work in the defense scheme proposed in [35]. In this paper, label sanitization (LS) bases its defense on the decision boundary of SVM, observing the remoteness of the poisoned samples, commending these samples to be re-labelled.
\\

Another way to detect outliers beforehand is seen in [58]. Notice that in our objectives detection and mitigation actions appear as the top priority. Data sanitization is meant to become a third type of defense to be used against an attacker that tries to tamper the training set, thereby, we would be more focus on the quality of the data before feeding it to the ML model of our choice.

\subsection{Third Approach: Defenses acting during inference phase}
Most of the works in literature do not consider defense systems acting during the inference phase once accounting for a solid defense active throughout the training phase. Possibly because such approach might not seem necessary since the hyperparameters of the model have been already set during training, guaranteeing that the integrity of the model is not compromised. For this reason, a defense approach acting during inference phase is not considered as a main point of interest in this present work.
\\

Once completed the training phase, the proposed defense mechanism should endow the target model with the necessary level of robustness to counteract the effects of poisoning data. Thereby the model should naturally reflect an improvement in performance during the inference phase, thus reducing considerably any potential drop in accuracy to a minimum if not negligible percent.

\section{Conclusions}

It is not new that ML model can be attacked by compromising the data needed in the training phase, jeopardizing entirely the decision-making process of classifiers. Several malware detection engines entailing ML applications have been developed in the past, this have proven not to be exempt of presenting vulnerabilities to be exploded by attackers. Attacks by label flipping with malicious intent until this day represent an important point of focus among researchers in the area. Moreover, there is an urgent need of a defense system to battle model-agnostic attacks, such as label-flipping. In contract, several works in the literature contemplate combating one type of attack in specific targeting one AI algorithm time of.

We have successfully proposed, developed and assessed two variations of the label flipping attack, this attacks have proven to be suited and tuned for a particular application based on a mobile ex-filtration data framework. Both attacks have demonstrated their capability to drastically reduce the overall accuracy and missclassification rate in one or both classes of different binary classifiers. Indeed this paper promises to focus a thesis research work into counteracting the effects of training data poisoning attacks, such as label flipping, on several types of machine learning models. 
\\

The aim and main contribution of this research centers around enhancing robustness of ML models against tampered training data to be used during re-training, continuous training of ML models in malware detection application is fundamental with many concerns regarding the reliability of this special types of binary classifiers. Then, once having understood the complexity of the proposed attack and its effects on the target model, we will be capable of detailing a more sophisticated algorithm based on the concept of described in this work; either as a defense technique that detects and later rejects the poisoning samples present in the training set or a more complex approach that entails a defense mechanism that can mitigate a drop in accuracy caused by training the model with poisoning data.  

\appendices
\section{ML Classifier Scripts}
Python code of the ML classifiers can be found in the following GitHub repository:\\
\underline{https://github.com/MiguelRamirezAguilar/PALM}
\section{Mobile Exfiltrarion Dataset}
The mobile exfiltration dataset employed in the training of the models covered in this work can be found in the following GitHub repository:\\
\underline{https://github.com/MiguelRamirezAguilar/PALM}

\begin{IEEEbiography}[{\includegraphics[width=1in,height=1.25in,clip,keepaspectratio]{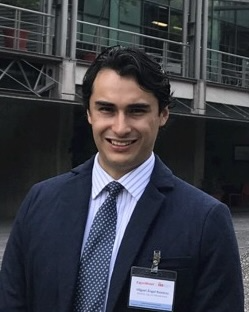}}]{Miguel Angel Ramirez Aguilar} Born in Mexico City, Mexico, 1994. Received his Bachelor’s Degree in Mechatronics Engineering from Universidad Nacional Autonoma de Mexico (UNAM), Mexico City, Mexico, in 2018, and is currently a MSc. student in Electrical and Computer Engineering specializing in Artificial Intelligence at Khalifa University, Abu Dhabi, United Arab Emirates.
Currently he is a Graduate Researcher at Khalifa University, United Arab Emirates; previously worked as a Computer-Aided Design engineer at Ford Motor Company, Mexico, R\&D Intern at SuitX “U.S. Robotics”, CA, USA, and as research assistant in robotics at Universidad Nacional Autonoma de Mexico, Mexico.
Published his thesis “Rediseño de dedo protésico,” Ptolomeo, 2020. [Online] Available: \underline {http://132.248.52.100:8080/xmlui/handle/132.248.52.100/17328} with main emphasis in robotics, optimization and prosthetics.
His current research interests are within the scope of Machine Learning oriented to Cybersecurity, Deep Learning and optimization algorithms. In addition, previous research experience related to the fields of control design, robotics and virtual instrumentation.
Mr. Miguel Angel Ramirez Aguilar, IEEE non-member.
\end{IEEEbiography}

\EOD

\end{document}